\documentclass{aa}
\usepackage{graphics}
\usepackage{natbib}
\bibpunct{(}{)}{;}{a}{}{,}
\newcommand{\WR}{\mathrm{WR}}
\newcommand{\Ostar}{\mathrm{O}}
\newcommand{\HeII}{\ion{He}{ii}}
\newcommand{\HeI}{\ion{He}{i}}
\newcommand{\arcdeg}{\circ}
\newcommand{\aap}{A\&A}
\newcommand{\aapr}{A\&AR}
\newcommand{\aaps}{A\&AS}
\newcommand{\apj}{ApJ}
\newcommand{\apjs}{ApJS}
\newcommand{\apss}{Ap\&SS}
\newcommand{\aj}{AJ}
\newcommand{\mnras}{MNRAS}
\newcommand{\pasp}{PASP}

\begin{document}
\title{Mass-Loss Rate Determination for the Massive Binary V444~Cyg using 3-D Monte-Carlo Simulations of Line and Polarization Variability}

\author{R. Kurosawa\inst{1}, D. J. Hillier\inst{1} and J. M. Pittard\inst{2}}

\institute{Department of Physics and Astronomy, University of Pittsburgh, 3941 O'Hara Street, Pittsburgh, PA 15260, USA \and Department of Physics \& Astronomy, The University of Leeds, Woodhouse Lane, Leeds LS2 9JT, UK}

\offprints{R.\ Kurosawa}

\mail{kurosawa@phyast.pitt.edu}

\date{Received DATE / Accepted DATE }

\abstract{A newly developed 3-D Monte Carlo model is used, in
conjunction with a multi-line non-LTE radiative transfer model, to
determine the mass-loss rate of the Wolf-Rayet (W-R) star in the
massive binary \object{V444~Cyg} (WN5+O6). This independent estimate
of mass-loss rate is attained by fitting the observed \HeI~\( 5876
\)~\AA~and \HeII~\( 5412 \)~\AA~line profiles, and the continuum light
curves of three Stokes parameters \( \left( I,\, Q,\, U\right) \) in
the \( V \) band simultaneously. The high accuracy of our
determination arises from the use of many observational constraints,
and the sensitivity of the continuum polarization to the mass-loss
rate. Our best fit model suggests that the mass-loss rate of the
system is \( \dot{M}_{\WR }=0.6\left( \pm 0.2\right) \times 10^{-5}\,
M_{\sun }\, \mathrm{yr}^{-1} \), and is independent of the assumed
distance to \object{V444~Cyg}. The fits did not allow a unique value
for the radius of the W-R star to be derived. The range of the volume
filling factor for the W-R star atmosphere is estimated to be in the
range of $0.050$ (for $R_{\WR}=5.0\, R_{\sun}$) to $0.075$ (for
$R_{\WR}=2.5\, R_{\sun}$). We also found that the blue-side of \HeI~\(
5876 \)~\AA~and \HeII~\( 5412 \)~\AA~lines at phase $0.8$ is
relatively unaffected by the emission from the wind-wind interaction
zone and the absorption by the O-star atmosphere; hence, the profiles
at this phase are suitable for spectral line fittings using a
spherical radiative transfer model. }

\authorrunning{Kurosawa et al.}

\titlerunning{Polarization and Mass-Loss Rate of V444~Cyg}

\maketitle

\section{Introduction}

\label{sec:Introduction}

The short-period (\( P=4.212 \) days, \citealp{khaliullin:1984})
eclipsing, massive binary V444~Cyg (WN5+O6 III-V) has been the subject
of extensive studies since its discovery \citep[e.g.,][]{wilson:1939, munch:1950, shore:1988, marchenko:1994, marchenko:1997}.
V444~Cyg exhibits variability in polarization, line strength and
X-ray flux as a function of orbital phase. The variability arises
from occultation of the photosphere, from perturbations induced in
the extended atmosphere of the Wolf-Rayet star (W-R) by the O star
and its wind, and from the wind-wind interaction region. Despite the
complexities, many authors \citep[][]{hamann:1992, stlouis:1993, marchenko:1994, cherepashchuk:1995, moffat:1996, marchenko:1997, stevens:1999}
have used this object to determine fundamental parameters of the W-R
star.

One of the most important and uncertain parameters in the stellar
evolution calculation of a massive star is the mass loss rate (\( \dot{M} \))
during the W-R star phase. Presently there are at least 6 different
methods of determining the mass-loss rate of a W-R star: 1.~dynamical
method using the change in the orbital period of binary \citep[e.g.,][]{khaliullin:1984},
2.~polarization variation method \citep{stlouis:1988}, 3.~X-ray spectra
synthesis method using a hydrodynamical model \citep[e.g.,][]{stevens:1996, pittard:2002},
4.~radio/IR continuum flux method \citep[e.g.][]{wright:1975}, 5.~radiative
transfer method \citep[e.g.,][]{hillier:1989, schmutz:1989} and 6.~photometric
variability method \citep[e.g.,][]{lamontagne:1996}. The mass-loss
rate estimated from methods 4 and 5 are, in general, about \( 2-3 \)
times higher than the values estimated from methods 1, 2 and 3. 

In the case of V444~Cyg, the mass-loss rate measured from the orbital
period change by \citet{khaliullin:1984}, \citet{underhill:1990} and
\citet{antokhin:1995} are \( \dot{M}_{\WR }=1.0\times 10^{-5}M_{\sun }\, \mathrm{yr}^{-1} \),
\( \dot{M}_{\WR }=0.4\times 10^{-5}M_{\sun }\, \mathrm{yr}^{-1} \)
and \( \dot{M}_{\WR }=0.7\times 10^{-5}M_{\sun }\, \mathrm{yr}^{-1} \)
respectively. (See also Table~\ref{tab:V444ParaPublished}.) \citet{prinja:1990}
found \( \dot{M}_{\WR }=2.4\times 10^{-5}\, M_{\sun }\mathrm{yr}^{-1} \)
from the free-free radio emission flux. By modeling the infrared lines, \citet{howarth:1992}
obtained \( \dot{M}_{\WR }=\left( 2.5-5\right) \times 10^{-5}\, M_{\sun }\, \mathrm{yr}^{-1} \).
More recently, \citet{nugis:1998} published the {}``clumping-corrected''
mass-loss rates for 37 W-R stars from the radio emission power and
the spectral index \( \left( \alpha =d\ln f_{\nu }/d\ln \nu \right)  \).
They obtained \( \dot{M}_{\WR }=0.92\times 10^{-5}\, M_{\sun }\, \mathrm{yr}^{-1} \)
for V444~Cyg. 

\citet{hamann:1992}, by using a non-LTE radiative transfer model with
a spherically expanding atmosphere, simultaneously fitted a set of
helium emission lines and the light curve of V444~Cyg. The mass-loss
rate estimated from their analysis is \( \dot{M}_{\WR }=1.26\times 10^{-5}\, M_{\sun }\, \mathrm{yr}^{-1} \).
Similar to Methods 4 and 5 mentioned above, the model assumes no clumps
in the stellar wind of the W-R star; hence, the mass-loss rate value
is most likely overestimated. An important conclusion of their work
is that the light curve is very sensitive to the inclination angle
which must be treated as a fitting parameter.

\citet[][hereafter STL1]{stlouis:1988} derived an analytical expression
for the mass-loss rate of a W-R star in a binary system, based on
the model of \citet{brown:1978} which predicts the continuum polarization
from a binary system as a function of orbital phase. The original
model of \citet{brown:1978} assumes point sources and a single scattering
atmosphere (optically thin). In the same paper, STL1 estimated \( \dot{M}_{\WR } \)
of V444~Cyg to be \( 0.9\times 10^{-5}\, M_{\sun }\, \mathrm{yr}^{-1} \)
assuming \( V_{\infty }=2500\, \mathrm{km}\, \mathrm{s}^{-1} \).
Later \citet[][hereafter STL2]{stlouis:1993} corrected this value
to \( \dot{M}_{\WR }=0.6\times 10^{-5}\, M_{\sun }\, \mathrm{yr}^{-1} \)
using the new estimate of \( V_{\infty }=1785\, \mathrm{km}\, \mathrm{s}^{-1} \)
\citep{prinja:1990}. STL2 also derived analytical expressions of Q
\& U Stokes parameters as a function of orbital phase \emph{especially
near the secondary eclipse} where the W-R is eclipsed by the O star.
The observed polarization curves were fitted with their model, and
the mass-loss rate of the W-R star in V444~Cyg was estimated (\( 0.75\times 10^{-5}\, M_{\sun }\, \mathrm{yr}^{-1} \)).
Although these models are very simple and potentially powerful, the
validity is still in question since the models assume that the W-R
star has a smooth wind and its atmosphere is optically thin. In addition,
the derived expression of \( \dot{M}_{\WR } \) in STL1 and \( Q \)
\& \( U \) in STL2 involves an integral which diverges. To avoid
the infinity, they fixed the lower radial integration limit to \( \sim 2R_{*} \).

Another important, but still uncertain parameter for V444~Cyg is
the luminosity ratio, \( q=L_{\WR }/L_{\Ostar } \). \citet{beals:1944}
spectroscopically estimated \( q \) for the wavelength range \( 4000-5000 \)\AA.
He obtained \( q=0.21 \) using the emission lines, and \( q=0.19 \)
using the absorption lines. For the same system, \citet{cherepashchuk:1995}
estimated the value of \( q \) in a similar manner as \citet{beals:1944},
but with some modification to the calculation of the equivalent width
of the emission/absorption lines. They found \( q=0.60\pm 0.06 \)
for \( \lambda \lambda  \)4000--6000; however, the values from different
lines show a large scatter, ranging from 0.36 to 0.80. From the two
principal emission lines used in these studies, they found \( q=0.51 \)
using \HeII~\( 5412 \)~\AA~and \( q=0.59 \) using \HeI~\( 5876 \)~\AA~(see
their Table~1). Lastly, \citet{hamann:1992} estimated \( q=3.1 \)
according to their models of the light curve and helium spectrum,
but the value is most likely incorrect since the effect of clumping
was not included in their model. 

The time dependent spectrum of the helium lines obtained by \citet{shore:1988},
\citet{underhill:1988}, \citet{marchenko:1994}, \citet{marchenko:1997}
and \citet{stevens:1999} shows very complex variability associated
with the orbital motion of the stars and the bow shock. Although the
terminal velocities of the O star (\( 2540\, \mathrm{km}\, \mathrm{s}^{-1} \): \citealp{underhill:1988})
and the W-R star (\( 1785\, \mathrm{km}\, \mathrm{s}^{-1} \): \citealp{prinja:1990})
are comparable, the W-R star has a much denser wind, c.f., \( \dot{M}_{\Ostar }\approx 0.6\times 10^{-6}\, M_{\sun }\, \mathrm{yr}^{-1} \)
\citep{leitherer:1988, marchenko:1997} and \( \dot{M}_{\WR }\approx \left( 0.4-1.26\right) \times 10^{-5}\, M_{\sun }\, \mathrm{yr}^{-1} \)
(Table~\ref{tab:V444ParaPublished}). The momentum carried by the
W-R wind is several times larger than that of the O star wind, and
the bow shock produced from the colliding stellar winds is folded
toward the O star. Further, the orbital speed of the binary is a significant
fraction of the terminal velocity of the W-R stellar wind (\( V_{\mathrm{orb}}/V_{\infty }\sim 1/4 \)
); hence, the shape of the bow shock is not expected to be cylindrically
symmetric around the axis joining the center of the W-R star and the
O star. In addition, the bow shock is not smooth, i.e., the shape
of the shock front is distorted by instabilities \citep[see][]{stevens:1994, gayley:1997, pittard:1998, pittard:1999}
--- the Kelvin-Helmholtz and the non-linear thin-shell instabilities
discussed in \citet{chandrasekhar:1961}, \citet{vishniac:1983}, \citet{vishniac:1994} and
\citet{blondin:1996}.

The first X-ray evidence of the colliding winds (CWs) in V444~Cyg
was found by \citet{moffat:1982} from the flux variability seen in
Einstein Observatory data. In addition, \citet{corcoran:1996} and
\citet{maeda:1999} found a similar X-ray variability using ROSAT and
ASCA respectively. Theoretical aspects of CW related X-ray variability
have been developed by, for example, \citet{luo:1990}, \citet{usov:1990}, \citet{usov:1992},
\citet{stevens:1992}, \citet{myasnikov:1993} and \citet{pittard:1997}.

In a previous paper \citep[][hereafter, Paper I]{kurosawa:2001a},
we developed a 3-D Monte-Carlo radiative transfer model which predicts
the level of continuum and line polarization produced by scattering
of light in a stellar atmosphere with arbitrary geometry. The model
can predict the variability features associated with orbital motion
of a binary system, including the polarization level, the flux level
and line profile shapes. The model can treat a finite sized stellar
disk, multiple scattering, line absorption of the continuum photons
and emission from multiple light sources. To achieve high precision
with a minimum of data points (i.e. to save computer memory), an 8-way
tree data structure created via a {}``cell-splitting'' method \citep[e.g,][]{wolf:1999}
is used in the model. This method is essential since the model assumes
no symmetry in the atmosphere. A logarithmic grid, which is commonly
used in spherical and axi-symmetric codes, is not readily implementable
in a model of arbitrary geometry.

The goal of this paper is to apply the 3-D Monte-Carlo model developed
in Paper I to V444~Cyg, and to interpret the observed variability
features (seen in the continuum polarization and line profile shape)
as a function of orbital phase. We estimate the mass-loss rate of
the W-R component in the binary, the orbital inclination, and the
luminosity ratio of the two stars, by fitting the observed \HeI~\( 5876 \)~\AA~and
\HeII~\( 5412 \)~\AA~line profiles, and the continuum light curves
of three Stokes parameters \( \left( \textrm{I},\, Q,\, U\right)  \)
in the \( V \) band consistently. Unlike the radio/IR/spectral methods
discussed earlier, the derived mass-loss rate determined by the polarization
method is insensitive to the amount of clumps in the stellar wind.
The continuum polarization level is proportional to the electron number
density while the thermal emissivity (of radio/IR) is proportional
to the square of the density. With a complicated density distribution,
a Monte Carlo simulation provides the only method to realistically
predict variability in polarization, continuum flux and line strength.
An original aim was also to obtain improved estimates of the stellar
parameters such as the radii of the W-R star and the O star, the monochromatic
luminosity ratio of the two stars and the orbital inclination angle.
Due to both model and observation uncertainties, this proved not to
be feasible. 

In \S\,\ref{sec:Analysis}, we discuss our models, examine the variability
of \HeI~\( 5876 \)~\AA~and \HeII~\( 5412 \)~\AA~lines, and
determine the mass-loss rate of the Wolf-Rayet star in V444~Cyg.
 The discussion on the results will be given in \S\,\ref{sec:Discussion},
and the conclusion in \S\,\ref{sec:Conclusion}.

\begin{table*}

\caption{Basic Parameters of V444 Cyg. \label{tab:V444ParaPublished}}

\vspace{0.3cm}
{\centering \begin{tabular}{ll}
 &
\\
\hline
\hline 
&
PUBLISHED VALUES\\
\hline
\( \dot{M}_{\WR } \)~ \( \left[ \times 10^{-5}\, M_{\odot }\, \mathrm{yr}^{-1}\right]  \)&
\( 0.6^{\left( a\right) } \), \( 1.0^{\left( b\right) } \), \( 0.4^{\left( c\right) } \),
\( \left( 0.8_{-0.3}^{+0.5}\right) ^{\left( d\right) } \), \( 0.7^{\left( v\right) } \),
\( 0.9^{\left( w\right) } \)\\
&
\( 2.4^{\left( e\right) } \), \( 2.5-5.0^{\left( f\right) } \),
\( 1.26^{\left( g\right) } \), \( 0.92^{\left( o\right) } \)\\
\( q\, \left[ =L_{\WR }\left( \lambda \right) /L_{\Ostar }\left( \lambda \right) \right]  \)&
\( 0.6\pm 0.06^{(h)} \) for \( \lambda \lambda 4000-6000 \)~\AA\\
&
\( \sim 0.2^{\left( i\right) } \) \( \lambda \lambda 4057-6563 \)~\AA\\
&
\( 0.25-0.35^{\left( j\right) } \) for \( \lambda \lambda 4400-10000 \)~\AA\\
\( R_{\mathrm{WR}}\, \left[ R_{\sun }\right]  \)&
\( 2.9^{\left( r\right) } \), \( <4^{\left( a\right) } \), \( 6.31^{\left( g\right) } \),
\( 3.1-5.2^{\left( s\right) } \)\\
\( R_{\mathrm{O}}\, \left[ R_{\sun }\right]  \)&
\( 8.5\pm 1^{\left( a\right) } \), \( 10.0^{\left( r\right) } \),
\( 3.76^{\left( g\right) } \), \( 8.4-9.3^{\left( s\right) } \)\\
\( i\, \left[ \mathrm{deg}.\right]  \)&
\( 82.8\pm 0.9^{\left( l\right) } \), \( 78.8\pm 0.5^{\left( m\right) } \),
\( 80.8\pm 1.6^{\left( a\right) } \), \( 83.47^{\left( g\right) } \)\\
\( M_{v}^{\left( *\right) } \) (O+W-R)&
\( -5.7^{\left( n\right) } \) , \( -4.2^{\left( g\right) } \)\\
\( m_{v} \) (O+W-R)&
\( 8.27^{\left( n\right) } \)\\
\( DM \)&
\( 11.01\pm 0.20^{\left( u\right) } \), \( 11.27-11.47^{\left( d\right) } \),
\( 9.3^{\left( g\right) } \)\\
Mass (O:W-R) \( \left[ M_{\sun }\right]  \)&
\( 25 \) : \( 10^{\left( p\right) } \), \( 27.9 \)\( \left( \pm 3.2\right)  \)
: \( 9.3\left( \pm 1.0\right) ^{\left( q\right) } \), \( 37.5 \)
: \( 11.3^{\left( t\right) } \)\\
\hline
\end{tabular}\par}
\vspace{0.3cm}

(\( a \))~\citet{stlouis:1993}, (\( b \))~\citet{khaliullin:1984},
(\( c \))~\citet{underhill:1990}, (\( d \))~\citet{marchenko:1997},
(\( e \))~\citet{prinja:1990}, (\( f \))~\citet{howarth:1992},
(\( g \))~\citet{hamann:1992}, (\( h \))~\citet{cherepashchuk:1995},
(\( i \))~\citet{beals:1944}, (\( j \))~\citet{kuhi:1968}, (\( k \))~\citet{cherepashchuk:1975},
(\( l \))~\citet{piirola:1988}, (\( m \))~\citet{robert:1990},
(\( n \))~\citet{lundstrom:1984}, (\( o \))~\citet{nugis:1998},
(\( p \))~\citet{munch:1950}, (\( q \))~\citet{marchenko:1994},
(\( r \))~\citet{cherepashchuk:1984}, (\( s \))~\citet{moffat:1996},
(\( t \))~\citet{underhill:1988}, (\( u \))~\citet{forbes:1992},
(\( v \))~\citet{antokhin:1995}, (\( w \))~\citet{rodrigues:1995}.

\( \left( *\right)  \) assumed the distance to be 1.7 kpc and \( E_{b-v}=0.68 \)
\citet{lundstrom:1984} -- They assumed the intrinsic color to be \( \left( b-v\right) _{\mathrm{o}}=-0.30 \).

\end{table*}


\section{Analysis}

\label{sec:Analysis}

The five most important parameters which we initially hoped to constrain
in our analysis were: the mass-loss rate of the W-R star (\( \dot{M}_{\WR } \)),
the luminosity of the W-R star (\( L_{\WR } \)), the radius of the
W-R star (\( R_{\WR } \)), the volume filling factor (\( f \)) and
the monochromatic luminosity ratio of the W-R and O stars, \( q\left( \lambda \right) =L_{\WR }\left( \lambda \right) /L_{\Ostar }\left( \lambda \right)  \),
at \( \lambda =5630 \)\AA. The volume filling factor is defined as
the fractional volume which contains material (the higher density
regions or clumps), and it controls the amount of clumps in the atmosphere
(see e.g., \citealp{abbott:1981}). It can also be thought of as the
fractional length along any random line of sight which contains clumps. 

Since the procedure of the model fitting is rather complicated, we
summarize the steps for determining the parameters below:

\begin{enumerate}
\item Understand the basic behavior of the variability seen in \HeI~\( 5876 \)~\AA\, and
\HeII~\( 5412 \)~\AA\, lines to determine at which binary phase
the spectrum is least affected by the bow shock and by the presence
of the O star atmospheric absorption. In other words, determine at
which phase the spectrum is modeled best by a spherical radiative
transfer model (CMFGEN).
\item At the phase chosen to be the best in Step 1, fit the observed helium
spectrum for different W-R star radii, and find the corresponding
value of \( q \). In general, \( L_{\WR }=2\times 10^{5}\, L_{\sun } \)
is used%
\footnote{This luminosity value is chosen from the mass-luminosity relation
of pure Helium stars given in \citet{meynet:1994}, and \( M_{WR}\approx 10\, M_{\sun } \)
\citep{marchenko:1994}.
}.
\item Fit the continuum I, Q, U light curves around the secondary eclipse
(\( \phi \approx 0.5 \)) to constrain the radius of the W-R star.
(\emph{\( q \) and \( R_{\WR } \) are constrained}.)
\item With \( R_{\WR } \) and \( q \) values determined in Step~3, re-fit
the helium spectrum with different values of \( \dot{M} \).
\item Compute the continuum \( I \), \( Q \), \( U \) light curves for
the \( R_{\WR } \) , \( q \) and \( \dot{M}_{\WR } \) values used
in Step~4, and fit only the \( I \) light curve by adjusting the
orbital inclination (\( i \)) and \( R_{\Ostar } \).
\item Now there are several models which fit the observed \( I \) light
curve and the helium spectrum simultaneously, but only a certain value
of \( \dot{M}_{\WR } \) will fit the \( Q \) and \( U \) light
curves; therefore, the value of \( \dot{M}_{\WR } \) is constrained.
(\emph{\( i \), \( R_{\Ostar } \) and \( \dot{M}_{\WR } \) are
constrained}.)
\end{enumerate}

\subsection{Line Variability and Geometric Configuration of the Binary}

\label{subsec:LineVarBowShockModel}

Since the 3-D Monte Carlo polarization model introduced in Paper I
is not a self-consistent model (i.e., the radiation does not modify
the opacity of the atmosphere), the opacity and the emissivity must
be supplied as the inputs of a model. There are three possible sources
of the input field: 1.~parametrized functions from a semi-analytical
model, 2.~output from a spherical non-LTE radiative transfer model
with some modifications, and 3.~output from a hydrodynamical calculation
in conjunction with method~2. Method~1 is useful for testing models,
but not suited for modeling a real object. In our analysis, method~2
is mainly used. 

The first step, in method~2, is to compute the object's spectrum
using a non-LTE radiative transfer code (CMFGEN) which includes thousands
of lines to treat the line-blanketing effect \citep{hillier:1998}.
For example, see \citet{hillier:1999} and \citet{herald:2001} for
a detailed discussion of W-R star spectrum models across a wide wavelength
range. CMFGEN provides the opacity and the emissivity of the W-R star
and the O star separately, and is designed for a single star with
a {}``spherically'' expanding atmosphere; hence, the output fields
(opacities etc.) are functions of radius only. The second step is
to compute the variability of the Stokes parameters (\( I \), \( Q \),
\( U \)) and the line profiles by using the 3-D Monte Carlo model
as a function of the binary phase.

The velocity field in a CMFGEN model is assumed to have the following
form \citep{hillier:1999}: \begin{equation}
\label{eq:modefiedBetaVelocity}
v\left( r\right) =\frac{V_{0}+V_{1}+V_{2}}{1+\left( V_{0}/V_{\mathrm{core}}-1\right) \, \exp \left[ -\left( r-R_{*}\right) /h_{\mathrm{eff}}\right] }
\end{equation}
where \( V_{1}=\left( V_{\infty }-V_{\mathrm{ext}}-V_{0}\right) \left( 1-R_{*}/r\right) ^{\beta _{1}} \),
\( V_{2}=V_{\mathrm{ext}}\left( 1-R_{*}/r\right) ^{\beta _{2}} \),
\( h_{\mathrm{eff}} \) is an isothermal effective scale height in
the inner atmosphere, \( V_{\infty } \) is the terminal velocity,
\( V_{\mathrm{core}} \) is the expansion velocity of the core, and
\( V_{\mathrm{ext}} \) \& \( \beta _{2} \) are the parameters for
the outer parts of the wind. \( \beta _{1} \) is equivalent to the
\( \beta  \) in the classic beta-velocity law, \( V\left( r\right) =V_{\infty }\left( 1-R_{*}/r\right) ^{\beta } \),
for the {}``inner part'' of the stellar wind. See \citet{najarro:1997}
and \citet{hillier:1999} for more explanations on this velocity law.
Although the line-driving of O-star winds is well understood, the
multiple-scattering of W-R winds complicates theoretical models, and
the standard CAK theory \citep*{castor:1975} does not apply for W-R
stars. 

\begin{figure}
\vspace{0.3cm}
{\centering \resizebox*{1\columnwidth}{!}{\includegraphics{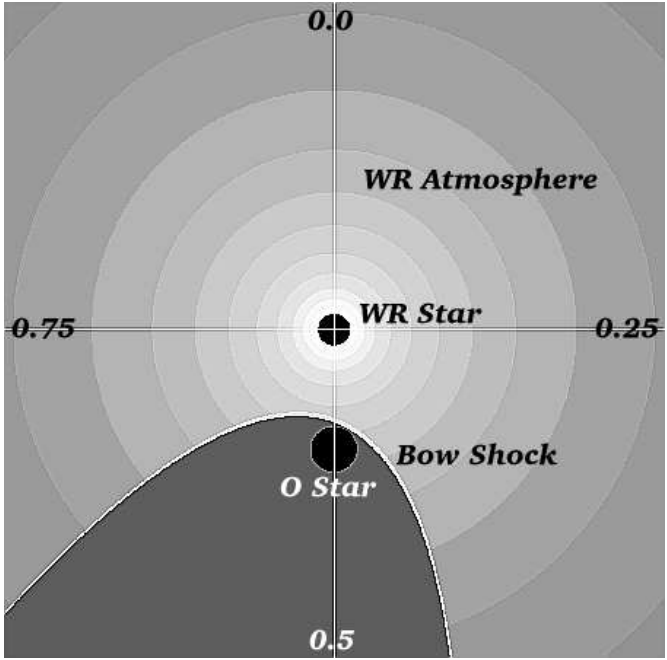}} \par}
\vspace{0.3cm}

\caption{\label{fig:ModelConfig} This figure illustrates the model
configurations of V444~Cyg. The W-R star is placed at the center
of the cubic boundary, and it is surrounded by a spherical atmosphere
which is consistent with the CMFGEN spectral model. The O star is
located below the W-R star in the diagram, and is covered by the tilted
paraboloid bow shock with a given thickness. The strong stellar wind
from the W-R star is interrupted by the shock front, and the density
behind the shock is assumed to be negligibly small for simplicity.
The directions of an observer at phase \( =0,\, 0.25,\, 0.5,\, 0.75 \)
are indicated at the top, bottom, left and right edges since the orbital
inclination is about \( 80^{\arcdeg } \) (almost edge-on view). }
\end{figure}

Fig.~\ref{fig:ModelConfig} shows the basic configuration of the model
geometry. The W-R star with its extended atmosphere is located at
the center. Surrounding the O star, the bow shock due to the CWs has
a paraboloidal shape, and points away from the W-R star since its
wind is much stronger than that of the O star. The head of the shock
front is tilted from the purely radial direction because of the rapid
orbital motion. Using the ratio of the terminal velocity of the W-R
component (\( V_{\infty }=1785\, \mathrm{km}\, \mathrm{s}^{-1} \),
\citealp{prinja:1990}) and the orbital speed (\( V_{\mathrm{orb}}\approx 460\, \mathrm{km}\, \mathrm{s}^{-1} \))%
\footnote{We simply estimated the value by assuming a circular orbit. If the
radial velocity measurements of \citet{marchenko:1994} and the orbital
inclination \( i=78.8^{\arcdeg } \) \citep{robert:1990} were used,
we would obtain \( V_{\mathrm{orb}}\approx 440\, \mathrm{km}\, \mathrm{s}^{-1} \). Consequently,
the tilt angle would be \( \delta \approx 14^{\arcdeg } \). 
}, the tilt angle of the bow shock region can be estimated as \( \delta \simeq \tan ^{-1}(V_{\mathrm{orb}}/V_{\infty })\approx 15^{\arcdeg } \).
Extensive discussions on the morphology of the bow shock can be found
in \citet{marchenko:1997}, including the issue of whether the W-R
wind is impacting onto the O star surface or not. \citep[See also the hydrodynamical calculation of][]{pittard:1999}.

At \( \phi  \) (orbital phase) \( =0 \), the W-R star is in front
of the O star, and an observer is looking at the system from the top
of Fig.~\ref{fig:ModelConfig} since the inclination angle \( i \)
is about \( 80^{\arcdeg } \) (almost edge-on view). At \( \phi =0.5 \),
the O star is in front of the W-R star. The directions of an observer
at \( \phi =0,\, 0.25,\, 0.5,\, 0.75 \) are indicated in the same
figure, and they infer the sense of the orbital motion (counter-clockwise
in the figure). The strong stellar wind from the W-R star is interrupted
by the shock front, and the density behind the shock is assumed to
be negligibly small for simplicity. In other words, we do not include
the O star wind explicitly, and we place the bow shock zone rather
artificially.

\begin{figure}
\vspace{0.3cm}
{\centering \resizebox*{1\columnwidth}{!}{\includegraphics{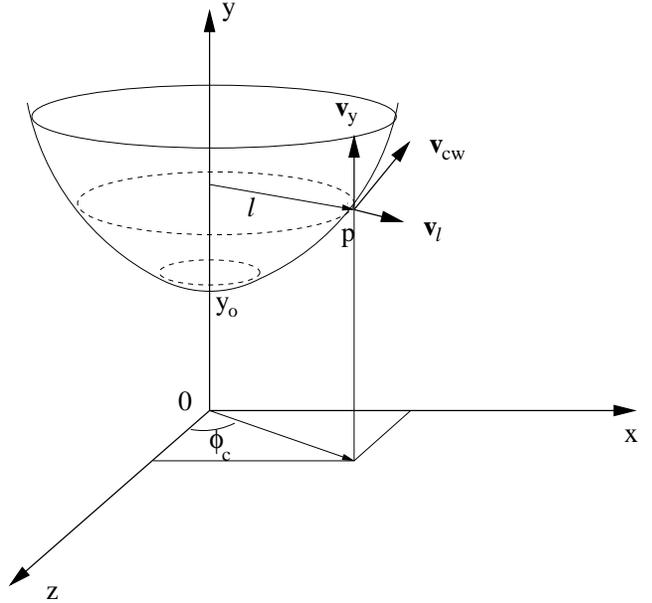}} \par}
\vspace{0.3cm}

\caption{\label{fig:CWmodel}This illustrates the simple geometric
model of the bow shock and the flow of the gas. The W-R star is located
at the origin (0), and the O star is located just behind the bow shock
on the y axis. The gas in the bow shock region is assumed to flow
tangentially to the surface of the paraboloid in the \( +y \) direction.
The gas velocity \( \mathbf{v}_{\mathrm{cw}} \) at a point \( p \)
on the surface is decomposed into \( v_{y} \) and \( v_{l} \), but
there is no azimuthal component (i.e., no rotation about the \( y \)
axis).}
\end{figure}

Next, we propose a simple model of the gas flow in the CW zone and
the geometry of the CW zone. We assume the shape of the CW zone to
be paraboloid%
\footnote{See \citet{huang:1982}, \citet{girard:1987} and \citet{canto:1996}
for the predicted geometry of the thin-shell shocks created by the
interaction of the two stellar winds in a highly radiative binary.
} with thickness \( d \). The center of the paraboloid is located
along the y-axis but displaced by \( y_{o} \) from the origin as
shown in Fig.~\ref{fig:CWmodel}. The surface of the paraboloid can
be written as:

\begin{equation}
\label{eq:paraboloid}
y=a_{y}l^{2}+y_{o}
\end{equation}
where \( l^{2}=x^{2}+z^{2} \). We define the bow shock region to
be the volume between this surface and the same surface displaced
by \( d \) (thickness) in \( +y \) direction.

The flow of the gas in the CW zone is assumed to be tangential to
the surface of the paraboloid; hence, there is no azimuthal component.
Then, the velocity (\( \mathbf{v}_{\mathrm{cw}} \)) of the gas flow
at point \( P \) in the CW zone can be written as:\begin{equation}
\label{eq:CWflow}
\mathbf{v}_{\mathrm{cw}}=v_{y}\hat{\mathbf{y}}+v_{l}\hat{\mathbf{l}}
\end{equation}
 where \( \hat{\mathbf{l}}=\cos \phi _{c}\hat{\mathbf{z}}+\sin \phi _{c}\hat{\mathbf{x}} \).
\( v_{y} \) and \( v_{l} \) are the velocity components which are
parallel and perpendicular to the y axis, respectively. \( a_{y} \)
is a parameter to control the asymptotic open angle of the bow shock.
Further, by taking the derivative of Eq.~\ref{eq:paraboloid} and
using the absence of an azimuthal component of \( \mathbf{v}_{\mathrm{cw}} \),
we obtain:

\begin{equation}
\label{eq:tangential}
\frac{dy}{dl}=2a_{y}l=\frac{v_{y}}{v_{l}}\, .
\end{equation}

If the magnitude of the velocity (\( \mathbf{v}_{\mathrm{cw}} \))
at point \( P \) is similar to that of the spherical flow (\( v_{r} \))
of the pre-shock W-R wind at \( P \) , i.e., \( v_{\mathrm{cw}}\approx v_{r} \),
then 

\begin{equation}
\label{eq:VrApproxmate}
v^{2}_{r}\approx v^{2}_{y}+v^{2}_{l}\, .
\end{equation}
 Since the flow of gas in the bow shock region is usually slower than
the surrounding flow according to the hydrodynamics model of \citet{pittard:1999},
we introduce a free parameter, and rewrite the equation above as:

\begin{equation}
\label{eq:VrParametrized}
v_{\mathrm{cw}}=s\, v_{r}=\left( v_{y}^{2}+v^{2}_{l}\right) ^{1/2}.
\end{equation}
 In general \( s<1 \), and is a function of the position (e.g., \( s=s\left( y\right)  \)).
However, in most of the analysis, \( s=1/2 \) is assumed for simplicity.

Using Eqs.~\ref{eq:tangential}, \ref{eq:VrApproxmate} and \ref{eq:VrParametrized},
the components of \( \mathbf{v}_{\mathrm{cw}} \) can be expressed,
in terms of the position \( \left( y,\, l\right)  \) on the paraboloid,
as \begin{equation}
\label{VyComponent}
v_{y}=\frac{2a_{y}\, l\, s\, v_{r}}{\left[ 1+\left( 2a_{y}l\right) ^{2}\right] ^{1/2}}\, ,
\end{equation}

and

\begin{equation}
\label{eq:VrhoComponent}
v_{l}=\frac{s\, v_{r}}{\left[ 1+\left( 2a_{y}\, l\right) ^{2}\right] ^{1/2}}\, .
\end{equation}
 These expressions are then corrected for the tilt of the bow shock
due to the fast orbital motion of the system, as mentioned earlier.
A more accurate velocity field in the bow shock region is needed to
properly model the line profile and the line polarization, but it
is not important for the continuum calculations. 

Considering the radial dependence of the density, we define: \begin{equation}
\label{eq:CWdensity02}
\rho _{\mathrm{shock}}=\rho _{\mathrm{o}}\left( \frac{y_{\mathrm{o}}}{r}\right) ^{n}
\end{equation}
 where \( r \) is the distance between the origin (\( 0 \)) and
a point \( p \) on the surface of the parabola, and \( y_{\mathrm{o}} \)
is the distance between the origin and the head of the paraboloid
as shown in Fig.~\ref{fig:CWmodel}. The index \( n \) is a scaling
parameter, and \( \rho _{\mathrm{o}} \) is the density at the head
of the parabola (at \( y=y_{\mathrm{o}} \)). This formulation is
similar to one in \citet{stevens:1999}. \( \rho _{\mathrm{o}} \)
is estimated from the amount of the excess emission seen in the \HeI~\( 5876 \)~\AA~line,
and is discussed in the later section (\S~\ref{subsubsec:BowshockEffect}).
In our analysis, we use the bow shock density expressed in Eq.~\ref{eq:CWdensity02}
with \( n=2 \), and the thickness (\( d \)) of the bow shock \( \approx 0.1\, R_{\Ostar } \),
estimated from \citet{pittard:1999}. In predicting the continuum polarization
level, the global distribution of the electron density is more important
than small structure. The column density or the electron scattering
opacity through the bow shock is related to the amount of gas in the
cavity which will be absent if there is no bow shock. 

\begin{figure}
{\centering \resizebox*{1\columnwidth}{!}{\rotatebox{270}{\includegraphics{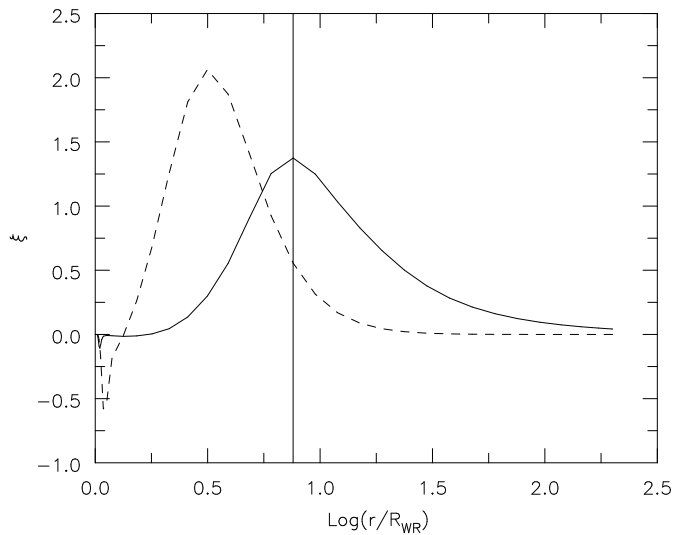}}} \par}

\caption{\label{fig:HeEmission}Illustration of the approximate location where \HeI~$5876$~\AA\,(solid) and \HeII~$5412$~\AA\, (dashed) lines originate according to a CMFGEN model of a typical WN5 star with $R_{\WR }=5\, R_{\sun }$. The amount of line emission originating in the interval $d\log \left( r/R_{\WR }\right)$ is proportional to $\xi \, d\log \left( r/R_{\WR }\right)$ where $r$ is the distance from the center of the W-R star. The solid vertical line at $\log \left( r/R_{\WR }\right) =0.88$ indicates the location of the O star assuming that the separation of system is $a=38\, R_{\sun }$ \citep{marchenko:1994}. According to this figure, the \HeII\, emission peaks at around $3\, R_{\WR }$ from the center of the W-R star, and the \HeI\, emission peaks around the location where the O star resides.}
\end{figure}

Some sensitivity of the line strengths and polarization may arise
from assuming whether the helium is singly or doubly ionized. \HeI~\( 5876 \)~\AA~and
\HeII~\( 5412 \)~\AA~lines are often referred to as {}``diagnostic
lines'' for modeling a WN star atmosphere \citep[e.g.,][]{hamann:1995}.
The latter is an optically thick line; hence, the absorption and scattering
terms in radiative transfer equations must be treated properly. The
gas in the bow shock is rapidly cooled by radiative processes, and
it could potentially cause a significant amount of absorption in the
\HeI~\( 5876 \)~\AA~line. Figure.~\ref{fig:HeEmission} shows
the approximate location where \HeI~\( 5876 \)~\AA~and \HeII~\( 5412 \)~\AA~lines
originate according to a CMFGEN model of a typical WN5 star with \( R_{\WR }=5\, R_{\sun } \).
The figure shows that the \HeII~emission peaks at around \( 3\, R_{\WR } \)
from the center of the W-R star, and the \HeI~emission peaks around
the location where the O star resides. This explains why the emission
from the CW zone is more clearly seen in the \HeI~line. In the following,
the variability of the \HeI~line due to the presence of the bow shock
will be examined.

\begin{figure}
{\centering \resizebox*{1\columnwidth}{!}{\includegraphics{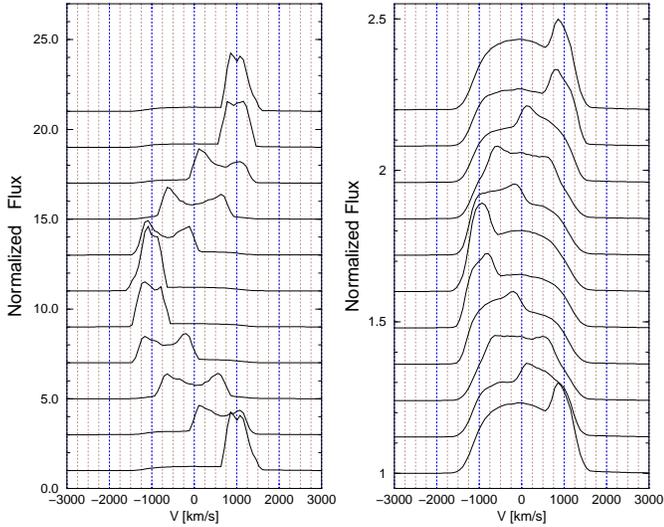}} \par}

\caption{\label{fig:CWemission} Left: The phase dependent emission from the
bow shock region on top of \HeI~\protect\( 5876\protect \)~\AA~line.
The emissivity of the bow shock is assigned to be unrealistically
strong in order to demonstrate its effect more clearly. Right: As
for the left figure, but the emissivity of the bow shock is reduced
by a factor of \protect\( \sim 10\protect \). This figure qualitatively
describes the variability seen in the observation of this line \citep[e.g., see][]{marchenko:1994}.
The profiles are plotted from phase \protect\( =0\protect \) (bottom)
to phase \protect\( =1\protect \) (top) in 0.1 phase steps. The profiles
are normalized to continuum, and shifted upward by 2 (right) and 0.12
(left) consecutively at each phase. In these models, the tilt angle
\protect\( \delta =15^{\arcdeg }\protect \) and \protect\( \mathrm{open}\, \mathrm{angle}=90^{\arcdeg }\protect \)
\citep{marchenko:1994} are used.}
\end{figure}

\subsubsection{Effect of the Colliding Winds }

As explained earlier, the emission from the bow shock region is prominent
for some lines \citep[see ][]{marchenko:1994}. In this section, the
expected effect of the bow shock emission on the \HeI~\( 5876 \)~\AA~line
will be demonstrated using the 3-D Monte Carlo model. \HeI~\( 5876 \)~\AA~is
primarily a recombination line; hence, to first order we assume the
line emissivity in the bow shock region is simply proportional to
the square of the density. The tilt angle \( \delta =15^{\arcdeg } \)
and the \( \mathrm{open}\, \mathrm{angle}=90^{\arcdeg } \) \citep{marchenko:1994}
are used for the geometry of the bow shock. The left graph in Fig.~\ref{fig:CWemission}
shows a sequence of \HeI~\( 5876 \)~\AA~emission lines as a function
of phase (\( 0-1 \) with steps of \( 0.1 \) from the bottom) arising
from a V444~Cyg model which includes an \emph{unrealistically strong}
bow shock emission. In fact, the line emission is dominated by the
bow shock emission, and the underlying line emission from the W-R
atmosphere is barely visible in this figure. This plot demonstrates
the basic behavior of the emission from the bow shock region. (See
also \citealt{luhrs:1997} and \citealt{bartzakos:2001} for similar
models and discussion.) At \( \phi =0.8-1.0 \), the bow shock is
pointing away from an observer and tilted to the light of sight (Fig.~\ref{fig:ModelConfig});
therefore, the emission peak appears on the red side of the flat-top
\HeI~emission line. The shape of the CW emission changes very rapidly
from \( \phi =0 \) to \( 0.1 \) because a rapid change in the velocity
component of the bow shock arms along the line of sight occurs near
\( \phi =0.1 \). The emission from the bow shock flattens out during
\( \phi =0.1-0.4 \) and \( \phi =0.7-0.8 \) since the bow shock
arms are almost perpendicular to the direction of the observer. At
\( \phi =0.4-0.5 \), the peak appears on the blue side since the
bow shock arms are pointing towards an observer. A sequence of \HeI~\( 5876 \)~\AA~line
profiles with a more realistic level of bow shock emission is shown
in the right panel of Fig.~\ref{fig:CWemission}. The figure shows
that the line strength increases around \( \phi =0 \) and \( 0.5 \)
because the continuum level drops during the primary and secondary
eclipses.

\subsubsection{O star Absorption Effect }

The observed spectrum of V444~Cyg shows absorption lines due to the
O star atmosphere on top of the broad emission lines from the W-R
atmosphere. The position of absorption shifts as the relative speed
of the stars changes. The distortion caused by this effect is prominent
for some lines at certain binary phases. Since the model does not
take into account the photospheric spectrum of the O star, the O star
absorption lines have been added to the CMFGEN model in an approximate
manner. In principal they could easily be incorporated into the Monte
Carlo method by changing the appropriate boundary conditions. 

The O star component of V444~Cyg has spectral type O5-O6.5 III-V
\citep{marchenko:1994}. The O star (O6) spectrum is modeled separately
by CMFGEN, and the effect of the rotational broadening of lines is
added to the spectrum. In this process, the rotational speed \( V_{\mathrm{rot}}\, \sin i=215\, (\pm 13)\, \mathrm{km}\, \mathrm{s}^{-1} \)
\citep{marchenko:1994} is used. 

\begin{figure}
{\centering \resizebox*{1\columnwidth}{!}{\includegraphics{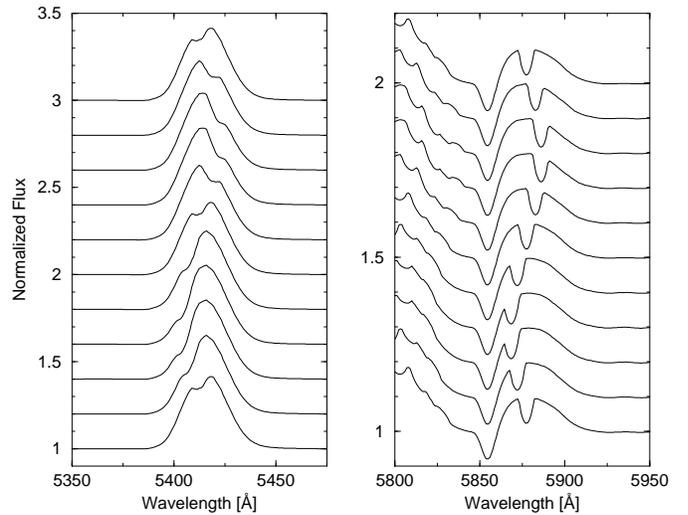}} \par}

\caption{\label{fig:OstarEffect}The figure illustrates how the O
star absorption affects the appearance of the \HeI~\( 5876 \)~\AA~and
\HeII~\( 5412 \)~\AA~lines from the W-R atmosphere as a function
of the orbital phase (\( \phi  \)). The spectrum of the W-R and the
O stars computed with the spherical model, CMFGEN, are combined using
the monochromatic luminosity ratio, \( q=L_{\WR }/L_{\Ostar }=0.45 \)
at \( \lambda =5630 \)~\AA. The figure does not include the effect
of the colliding wind for clarity. The profiles are plotted from \( \phi =0 \)
(bottom) to \( \phi =1 \) (top) with 0.1 phase steps.}
\end{figure}

Fig.~\ref{fig:OstarEffect} shows the combined spectrum of the W-R
and the O star computed by CMFGEN as a function of the orbital phase.
The sinusoidal motion of the O star absorption component is clearly
seen in both lines. To examine only the effect of the O star, \emph{the
emission from the CW zone is not included in the model spectra shown
in this figure}. A specific luminosity ratio (\( q=L_{\WR }/L_{\Ostar }=0.45 \)
at \( \lambda =5630 \)~\AA) was used when the two spectra were combined.

\subsubsection{Combined Effects Compared with Observation}

\begin{figure*}

{\centering \resizebox*{2\columnwidth}{!}{\includegraphics{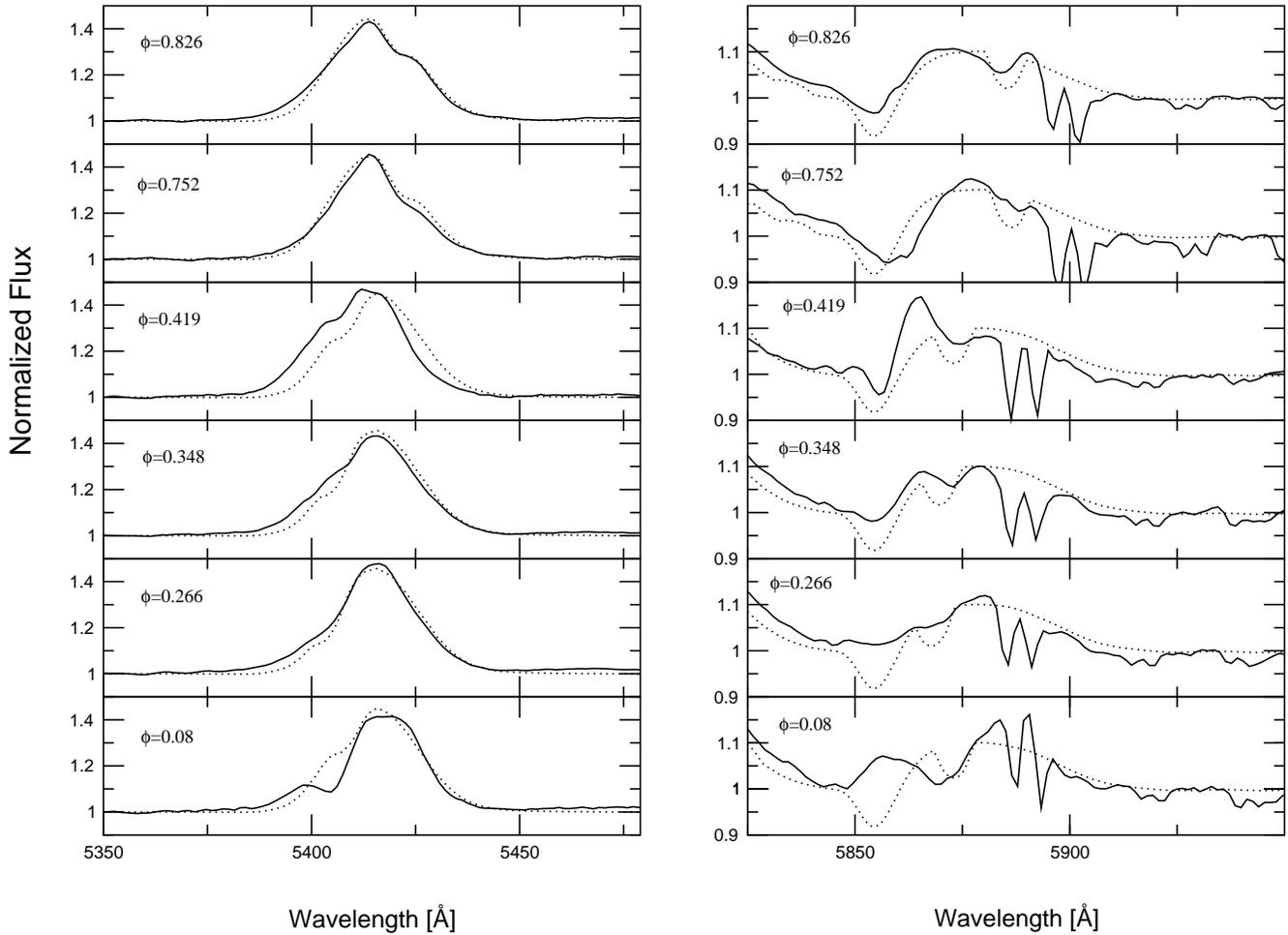}} \par}

\caption{\label{fig:HeFits}The figure shows one of the best fit CMFGEN
models (Dotted) at 6 different binary phases. The observational data
(Solid) were provided by Marchenko \citetext{priv. comm.}. The O star
model spectrum (O6V) is added to include the effect of the atmospheric
absorption effect. The monochromatic luminosity ratio \( q=L_{\WR }/L_{\Ostar }=0.45 \)
at \( \lambda =5630 \) \AA~ is used when the W-R star model spectrum
is combined with the O star model spectrum. In general, the fits are
good at all phases shown, especially at \( \phi \approx 0.3 \) and
\( \phi \approx 0.8 \). Two narrow \ion{Na}{i} interstellar absorption
lines are present in the red side of observed \HeI~\( 5876 \)~\AA~lines.
The emission from the bow shock is not included in the model shown
here.}
\end{figure*}

In order to use CMFGEN for the model fitting, we must choose a binary
phase when the lines are least affected by the bow shock emission
and the O star absorption. Fig.~\ref{fig:OstarEffect} shows that
the absorption by the O star does not affect the overall shape of
\HeI~\( 5876 \)~\AA, but it does affect the shape of \HeII~\( 5412 \)~\AA.
At \( \phi \approx 0.2 \) and \( \phi \approx 0.7 \), the \HeI~line
shape is least affected by the O star, but at these phases the emission
from the bow shock region is spread over the \HeI~line as seen in
Fig.~\ref{fig:CWemission}. As a result, the line flux level is likely
slightly higher than that of the spherical model. On the other hand,
at \( \phi \approx 0 \) and \( \phi \approx 0.5 \), the emission
from the bow shock region is well concentrated near the blue or the
red edge of the \HeI~line, but at these phases, the stars are eclipsed;
hence, the line strength will be underestimated in a simple spherical
model (CMFGEN). Meanwhile at \( \phi \approx 0.3 \) and \( \phi \approx 0.8 \),
the bow shock emission is concentrated only on either the blue or
the red side of the \HeI~line, and the O star absorption affects
the same half side of the line profile. Unfortunately, at \( \phi \approx 0.3 \),
the O star absorption is on the red side of the helium line, and the
bow shock emission is on the blue side of the line. So the line is
contaminated on both sides. On the other hand at \( \phi \approx 0.8 \),
both the O star absorption and the emission from CW are on the red
side of a line; hence we can reliably use the blue part of \HeI~\( 5876 \)~\AA~and \HeII~\( 5412 \)~\AA~for
the model fitting. Figure~\ref{fig:HeFits} shows the sample CMFGEN
model spectrum compared with the observed spectrum of Marchenko \citetext{priv. comm}
at six different phases. The agreement between the model and the observation
at \( \phi \approx 0.8 \) is excellent. Near the eclipsing phases
(\( \phi \approx 0 \) and \( 0.5 \)), the spherical model CMFGEN
does not fit the observation well since the contribution from the
bow shock is not included in the model. The two narrow \ion{Na}{i} interstellar
absorption lines are present in the red side of observed \HeI~\( 5876 \)~\AA~lines.

\subsection{Mass-Loss Rate Estimation of the W-R Component }

\label{subsec:MassLoss}

The idea behind the mass-loss estimation performed here is very similar
to the one in STL1 and STL2. We assume that there is a correspondence
between the amount of continuum polarization and the mass-loss rate
of the W-R component. This is supported by the analytic expression
of \( \dot{M} \) in terms of \( A_{p} \) (a semi-major axis of polarization
on Q-U plane) derived by STL1, using the basic results from the classic
paper on binary polarization by \citet{brown:1978}.

\subsubsection{Investigation of the W-R Radius}

\begin{figure}
{\centering \resizebox*{1\columnwidth}{!}{\includegraphics{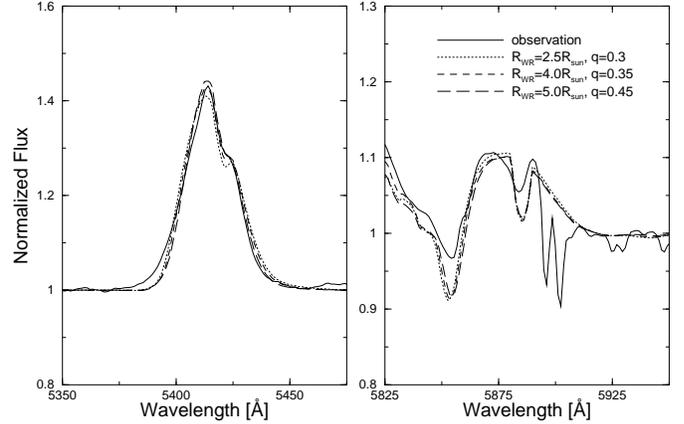}} \par}

\caption{\label{fig:BestFits} Left: the model fits of \HeII~\( 5412 \)~\AA~for
three different radii (\( 2.5 \), \( 4.0 \) and \( 5.0\, R_{\sun } \))
of the W-R star. Right: same as the left, but for \HeI~\( 5876 \)~\AA.
The profile shapes of both lines fit the observation well for a relatively
wide range of the W-R star radius. When the O star spectrum is added
to the W-R spectrum with an appropriate value of \( q \), the profiles
with different values of \( R_{\WR } \) are not significantly different
from each other. The values of \( q \) used are 0.30, 0.35 and 0.45
for \( R_{\WR }= \)2.5, 4.0 and \( 5.0\, R_{\sun } \) models respectively. The
corresponding volume filling factors of the models are 0.08, 0.04,
0.05.}
\end{figure}

For an assumed mass-loss rate, \( \dot{M} \), of the W-R star, there
are several CMFGEN models which can fit the observational spectrum
of \HeII~\( 5412 \)~\AA~and \HeI~\( 5876 \)~\AA~simultaneously.
As discussed in the previous section, the spectrum at \( \phi \approx 0.8 \)
is used in the model fitting since the contamination from the O star
atmospheric absorption and the CW emission is expected to be small
at this orbital phase. Fig.~\ref{fig:BestFits} shows the fits of
the observed spectrum at \( \phi =0.826 \) for different values of
\( R_{\WR } \) (=\( 2.5 \), \( 4.0 \), \( 5.0\, R_{\sun } \))%
\footnote{The different solutions correspond to different assumed distances
for V444 Cyg. The W-R radius (\( R_{\WR } \)) used in this paper
corresponds to the inner boundary of the model atmosphere where the
Rosseland optical depth \( \tau _{\mathrm{R}}\approx 30 \). 
} with the mass-loss rate of the W-R star fixed (\( \dot{M}_{\WR }=0.6\times 10^{-5}\, M_{\sun }\, \mathrm{yr}^{-1} \)).
The corresponding volume filling factors of the models are \( 0.08 \),
\( 0.04 \), \( 0.05 \). All three models fit the observation very
well; therefore, the spectral solutions are degenerate. The \HeII~profiles
with different values of \( R_{WR} \) are not significantly different
from each other when the O star spectrum is added to the W-R spectrum
with an appropriate value of \( q \) (\( =0.30,\, 0.35,\, 0.45 \)
for \( R_{\WR }=2.5,\, 4.0,\, 5.0\, R_{\sun } \) models respectively).
In the case of a single W-R star spectrum, a small but noticeable
difference in the profile shape of the \HeII~line, usually becoming
more triangular for a larger \( R_{\WR } \), should be present. Unfortunately,
it is hard to see this effect in the V444~Cyg model because the continuum
flux of the O star component is stronger (\( q\approx 0.5 \)); hence,
the line strength is weakened. 

\begin{figure}
{\centering \resizebox*{1\columnwidth}{!}{\includegraphics{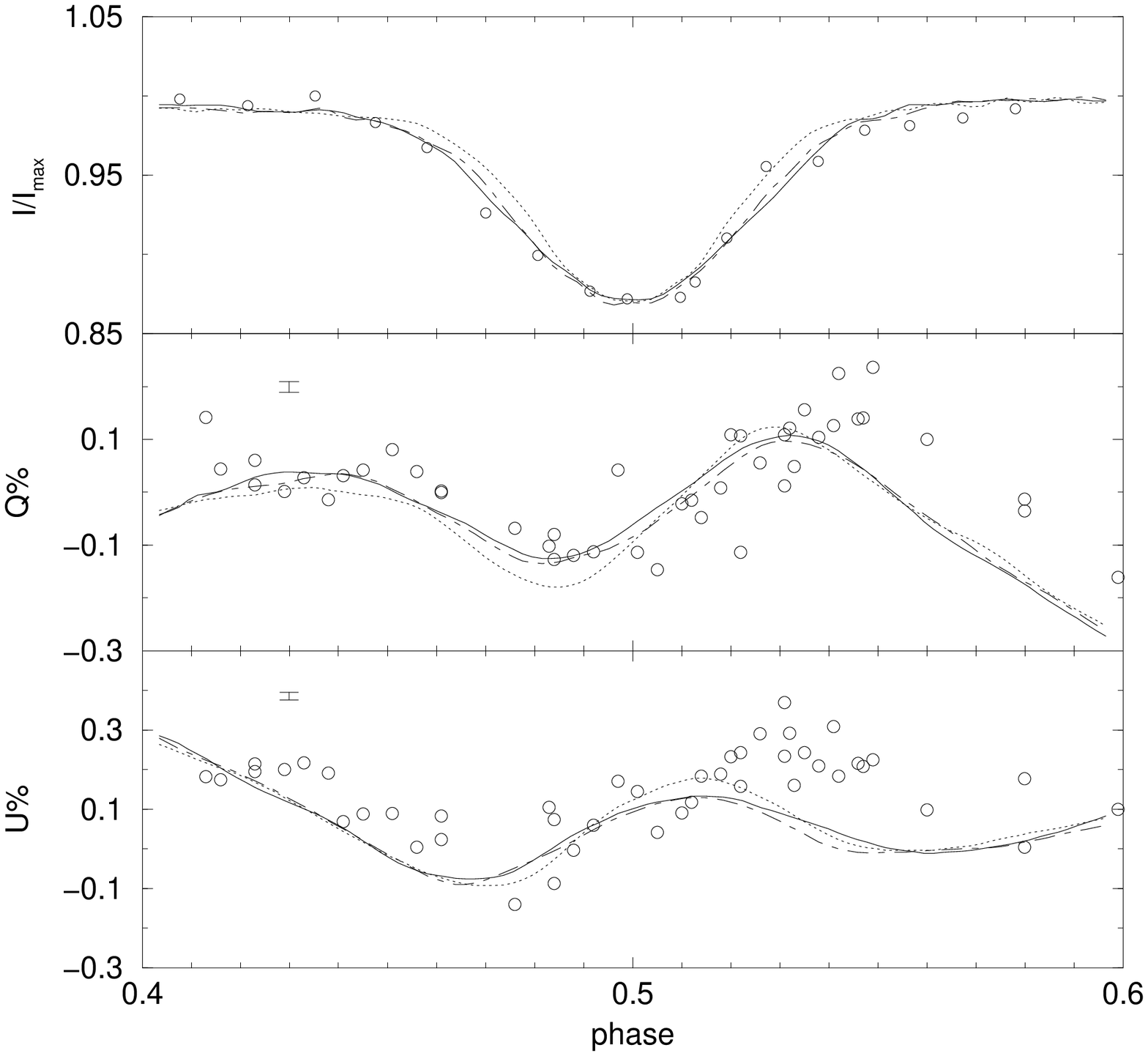}} \par}

\caption{\label{fig:RwrFits} The continuum flux, Q and U polarization
near the secondary eclipse (where the O star is in front of the W-R
star) are computed by the 3-D Monte-Carlo model of \citet{kurosawa:2001a} for
the three spectral models shown in Fig.~\ref{fig:BestFits}. Circles:
Observation, Dot: \( R_{\WR }=2.5\, R_{\sun } \) \& \( q=0.30 \)
model, Dash-Dot: \( R_{\WR }=4.0\, R_{\sun } \) \& \( q=0.35 \)
model, Solid: \( R_{\WR }=5.0\, R_{\sun } \) \& \( q=0.45 \) model.
The typical size of the error in model calculations of \( Q \) and
\( U \) is indicated near the upper left corner of the middle and
the bottom plots. The optical light curve data and optical polarization
data (V band) are from \citet{kron:1943} and STL2 respectively.}
\end{figure}

In an attempt to remove this degeneracy, the light curves (\( I \),
\( Q \), \( U \)) are computed by the 3-D Monte-Carlo model \citep{kurosawa:2001a} for
the spectral models shown in Fig.~\ref{fig:BestFits} for fixed \( R_{\Ostar } \)
and \( \dot{M}_{\WR } \) values. First, models without the bow shock
(Figs.~\ref{fig:RwrFits} -- \ref{fig:FinalFits2}) are considered
for clarity. The effect of the bow shock on the light curve will be
examined later. A small difference is expected to be seen (c.f., STL2)
in the shape of the polarization curve near the secondary eclipse
(\( \phi \approx 0.5 \), i.e., when the W-R star is behind the O
star).

The results, using \( R_{\Ostar }=7.2\, R_{\sun } \) and \( \dot{M}_{\WR }=0.6\times 10^{-5}\, M_{\sun }\, \mathrm{yr}^{-1} \),
are shown in Fig.~\ref{fig:RwrFits}. The orbital inclination angles
used for the fits are \( i=78.8^{\arcdeg } \), \( 78.5^{\arcdeg } \)
and \( 78.0^{\arcdeg } \) for the models with \( R_{\WR }=2.5\, R_{\sun } \),
\( 4.0\, R_{\sun } \) and \( 5.5\, R_{\sun } \) respectively. The
inclination angles are consistent with \( i=78^{\arcdeg }\pm 1^{\arcdeg } \)
\citep{cherepashchuk:1975} obtained from the light curve solution. STL2
estimated the inclination angle by fitting the double sine curves
of \( Q \) and \( U \) polarization components with the formula
given by \citet{brown:1978}. Their result is \( i=80.8^{\arcdeg }\pm 1.6^{\arcdeg } \).
In Fig.~\ref{fig:RwrFits}, the three models fit the observed \( I \)
light curve moderately well. Although not shown here, all three models
also fit well the observed \( I \) light curve outside of the phase
range shown in Fig.~\ref{fig:RwrFits}. The deviations from perfect
anti-symmetry around \( \phi =0.5 \) seen in the observational \( I \)
light curve and the anomaly seen around \( \phi \approx 0.53-0.54 \)
in the U observational light curve are not well fitted, and may be
related to the presence of the bow shock in the system.

The width of the \( I \) light curve around \( \phi =0.5 \) is slightly
wider for the model with the larger W-R star radius. The \( R_{\WR }=5.0\, R_{\sun } \)
model fits the overall shape of the light curve better than the two
other models with smaller radii. However, the width of the secondary
eclipse depends also on the mass-loss rate of the W-R star, and we
discuss this dependency later. Unfortunately, no significant difference
in the Q and U light curves are seen for different values of W-R star
radii, unlike that found by STL2. In fact, the difference in Q and
U light curves seen by STL2 for different \( R_{\WR } \) (see their
Fig.~8) is not purely due to changing \( R_{\WR } \), being also
coupled with \( \dot{M}_{\WR } \) and the adopted inclination. 

In this analysis, we were not able to constrain the value of \( R_{\WR } \).
Models with a wide range of \( R_{\WR }=2.5-5.0\, R_{\sun } \) can
fit the observed \( I \), \( Q \) and \( U \) light curves reasonably
well, and there is no significant difference among them as long as
appropriate values of the orbital inclination angle and the O star
radius are chosen for each model. Despite this uncertainty in the
W-R star radius, we continue to derive the mass-loss rate of the W-R
star.

\subsubsection{Determination of the W-R Mass-Loss Rate}

\begin{figure}
{\centering \resizebox*{1\columnwidth}{!}{\includegraphics{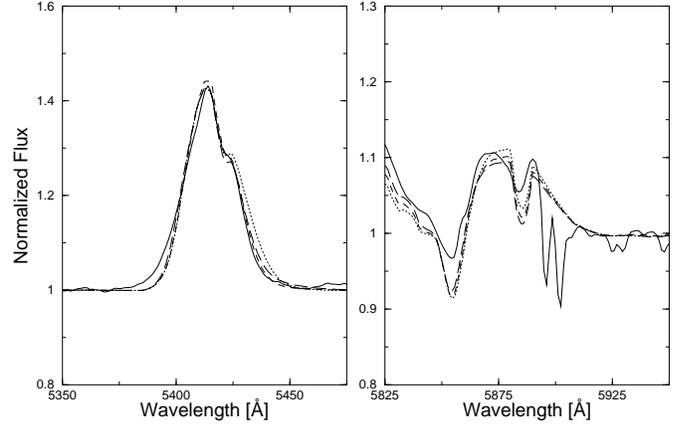}} \par}

\caption{\label{fig:5RsBestFits}The observed \HeII~\( 5412 \)~\AA~(left)
and \HeI~\( 5876 \)~\AA~(right) are fitted with the models for
different values of the W-R star mass-loss rate. \( R_{\WR }=5.0\, R_{\sun } \)
is used for all the models shown here. Solid: observation, Dot: Model~A
(\( \dot{M}_{\WR }=0.3\times 10^{-5}\, M_{\sun }\, \mathrm{yr}^{-1} \)),
Dash: Model~B (\( \dot{M}_{\WR }=0.6\times 10^{-5}\, M_{\sun }\, \mathrm{yr}^{-1} \)),
Long Dash: Model~C (\( \dot{M}_{\WR }=1.2\times 10^{-5}\, M_{\sun }\, \mathrm{yr}^{-1} \)). See
Table.~\ref{tab:ModelParameters} for other parameters used. The profiles
from all three models are very similar to each other, and their fits
to the observation are reasonable. It is impossible to distinguish
between the three models. }
\end{figure}

\begin{figure*}
{\centering \resizebox*{2\columnwidth}{!}{\includegraphics{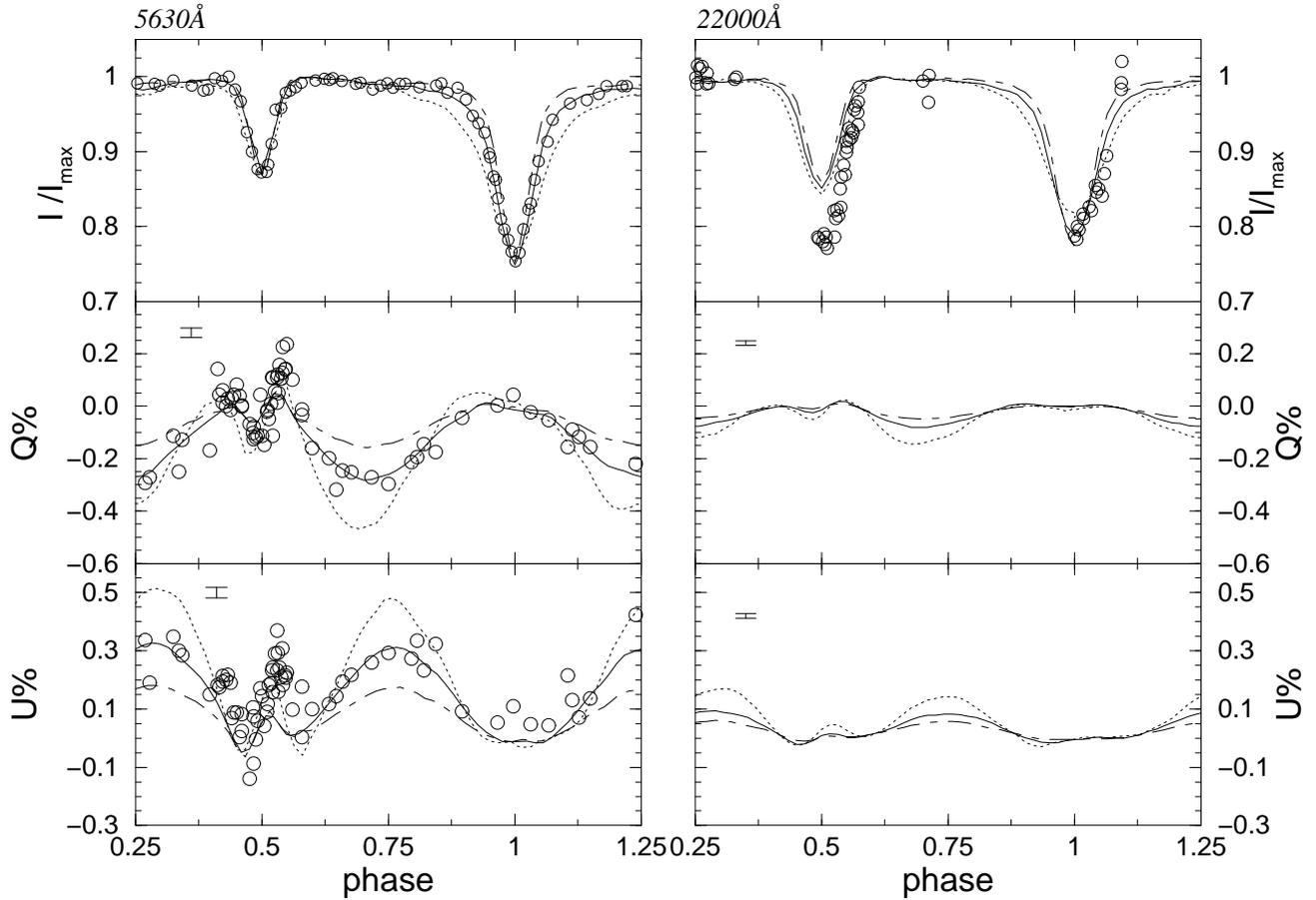}} \par}

\caption{\label{fig:FinalFits} For each model shown in Fig.~\ref{fig:5RsBestFits},
the polarization and flux variations are computed. The three plots
on the left are the relative flux, \( Q \) and \( U \) polarization
at \( \lambda =5630 \)\AA~as a function of the binary phase. The
three plots on the right side are the same as those on the left, but
computed at \( \lambda =2.2\, \mu \mathrm{m} \). Circle: observation,
Dash Dot: Model~A (\( \dot{M}_{\WR }=0.3\times 10^{-5}\, M_{\sun }\, \mathrm{yr}^{-1} \)),
Solid: Model~B (\( \dot{M}_{\WR }=0.6\times 10^{-5}\, M_{\sun }\, \mathrm{yr}^{-1} \)),
Dot: Model~C (\( \dot{M}_{\WR }=1.2\times 10^{-5}\, M_{\sun }\, \mathrm{yr}^{-1} \)).
The amplitude of the observed polarization curves is not consistent
with either Model~A or Model~C. On the other hand, Model~B (\( \dot{M}_{\WR }=0.6\times 10^{-5}\, \dot{M}_{\sun }\, \mathrm{yr}^{-1} \))
fits the observation very well. The optical light curve data and optical
polarization data are from \citet{kron:1943} and STL2 respectively.
The observed infrared light curve is from \citet{hartmann:1978}.}
\end{figure*}

Next, the observed helium spectrum is modeled for different values
of \( \dot{M}_{\WR } \) by separately adjusting the values of \( f \)
and \( q \) for \( R_{\WR }=5.0 \) and \( 2.5\, R_{\sun } \). Fig.~\ref{fig:5RsBestFits}
shows the results for \( \dot{M}_{\WR }=0.3 \), \( 0.6 \) and \( 1.2\times 10^{-5} \)\( M_{\sun }\, \mathrm{yr}^{-1} \)
with the fixed radius of the W-R star: \( R_{\WR }=5.0\, R_{\sun } \).
No significant difference in the profile shapes is seen in these models.
To remove this degeneracy, once again, the polarization curves were
computed; however, this time the amplitude of the {}``double-sine''
shaped polarization curves were fitted. The polarization curves for
each value of \( \dot{M}_{\WR } \) corresponding to the spectral
models in Fig.~\ref{fig:5RsBestFits} are shown in Fig.~\ref{fig:FinalFits}
along with the light curves at \( \lambda =5630 \)~\AA. The model
parameters used for the fits in Fig.~\ref{fig:FinalFits} are summarized
in Table~\ref{tab:ModelParameters}. The figure clearly shows that
the model with \( \dot{M}_{\WR }=1.2\times 10^{-5}\, M_{\sun }\, \mathrm{yr}^{-1} \)
overestimates and the one with \( \dot{M}_{\WR }=0.3\times 10^{-5}\, \dot{M}_{\sun }\, \mathrm{yr}^{-1} \)
underestimates the amplitude of the observed \( Q \) and \( U \)
curves, while \( \dot{M}_{\WR }=0.6\times 10^{-5}\, \dot{M}_{\sun }\, \mathrm{yr}^{-1} \)
model is consistent with the observations. The width of the \( I \)
light curve for \( \lambda =5630 \)~\AA~ around the primary eclipse
(\( \phi =0.0 \)) with \( \dot{M}_{\WR }=0.3\times 10^{-5}\, M_{\sun }\, \mathrm{yr}^{-1} \)
is slightly narrower than the observation, and that with \( \dot{M}_{\WR }=1.2\times 10^{-5}\, M_{\sun }\, \mathrm{yr}^{-1} \)
is slightly wider than the observation. The observed \( I \) light
curve at \( \lambda =5630 \)~\AA~ is fitted with the \( \dot{M}_{\WR }=0.6\times 10^{-5}\, M_{\sun }\, \mathrm{yr}^{-1} \)
model very well.

\begin{figure}
{\centering \resizebox*{1\columnwidth}{!}{\includegraphics{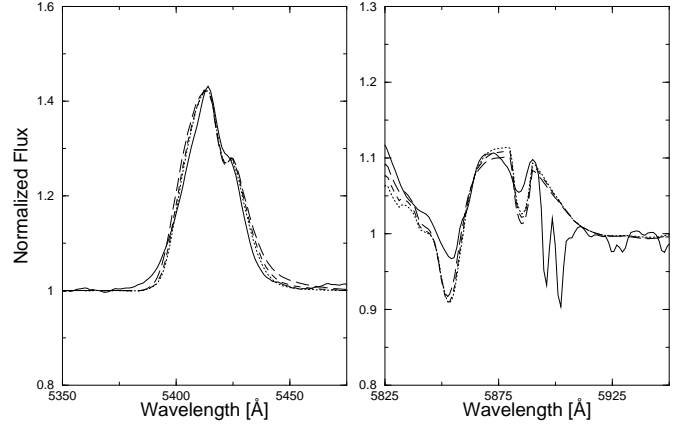}} \par}

\caption{\label{fig:2halfRsBestFits} Same as Fig.~\ref{fig:5RsBestFits},
but with \( R_{\WR }=2.5R_{\sun } \). Solid: observation, Dot: Model
D (\( \dot{M}_{\WR }=0.3\times 10^{-5}\, M_{\sun }\, \mathrm{yr}^{-1} \)),
Dash: Model E (\( \dot{M}_{\WR }=0.6\times 10^{-5}\, M_{\sun }\, \mathrm{yr}^{-1} \)),
Long-Dash: Model F (\( \dot{M}_{\WR }=1.2\times 10^{-5}\, M_{\sun }\, \mathrm{yr}^{-1} \)).
See Table~\ref{tab:ModelParameters} for other parameters used.}
\end{figure}

\begin{figure*}
{\centering \resizebox*{2\columnwidth}{!}{\includegraphics{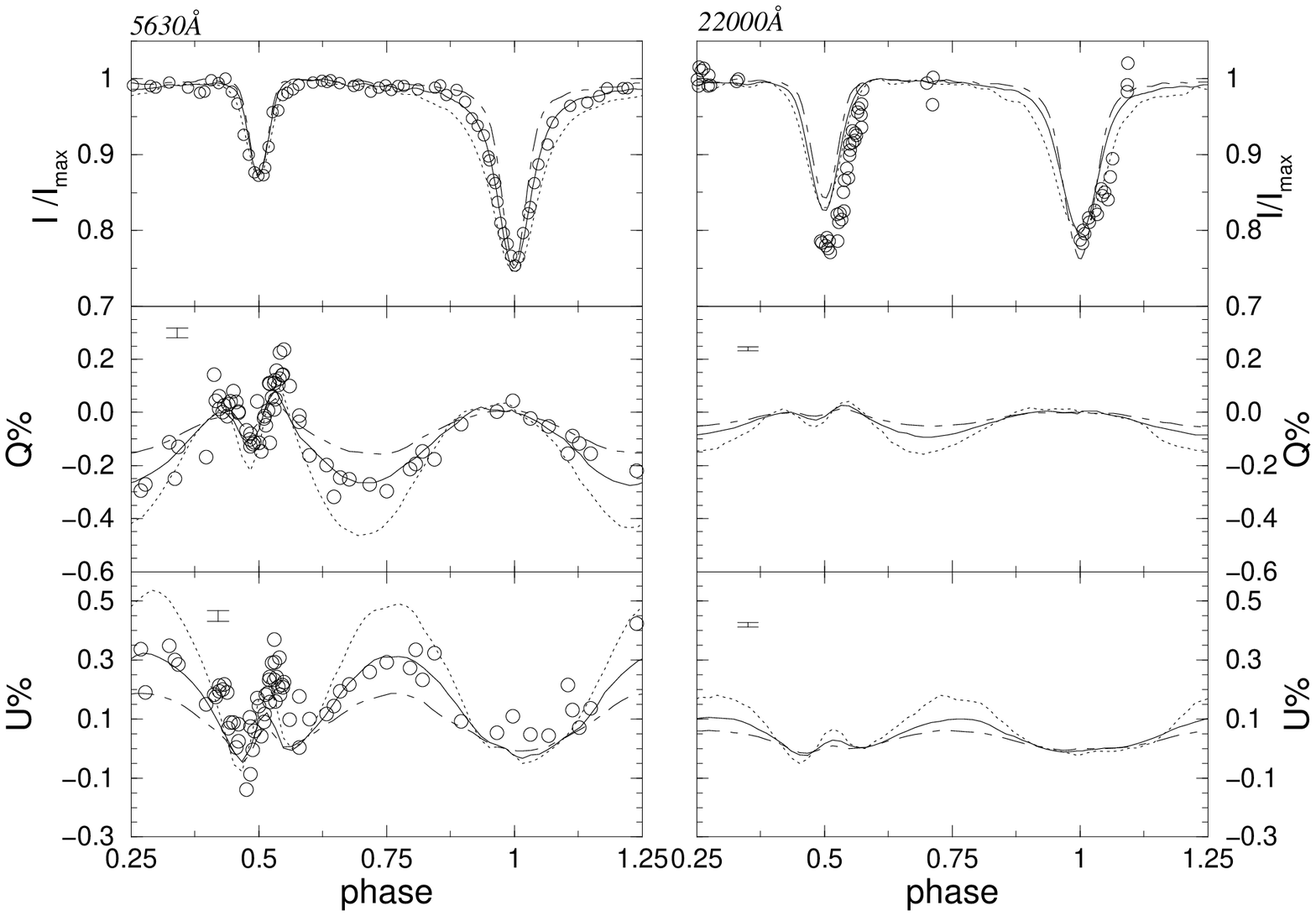}} \par}

\caption{\label{fig:FinalFits2}Same as Fig.~\ref{fig:FinalFits},
but the models are calculated for \( R_{\WR }=2.5\, R_{\sun } \).
The spectral models corresponding to these models are shown in Fig.~\ref{fig:2halfRsBestFits}.
Circle: observation, Dash-Dot: Model D (\( \dot{M}_{\WR }=0.3\times 10^{-5}\, M_{\sun }\, \mathrm{yr}^{-1} \)),
Solid: Model E (\( \dot{M}_{\WR }=0.6\times 10^{-5}\, M_{\sun }\, \mathrm{yr}^{-1} \)),
Dot: Model F (\( \dot{M}_{\WR }=1.2\times 10^{-5}\, M_{\sun }\, \mathrm{yr}^{-1} \)).
Model~D (\( \dot{M}_{\WR }=1.2\times 10^{-5}\, M_{\sun }\, \mathrm{yr}^{-1} \))
and Model~F (\( \dot{M}_{\WR }=0.3\times 10^{-5}\, \dot{M}_{\sun }\, \mathrm{yr}^{-1} \))
do not fit the amplitudes of the observed polarization curves. On
the other hand, Model~E (\( \dot{M}_{\WR }=0.6\times 10^{-5}\, \dot{M}_{\sun }\, \mathrm{yr}^{-1} \))
fits the observation very well.}
\end{figure*}

Since there is a large uncertainty in the \( R_{\WR } \) value, the
results of similar calculations with \( R_{\WR }=2.5\, R_{\sun } \)
are shown in Figs.~\ref{fig:2halfRsBestFits} and \ref{fig:FinalFits2}.
They are very similar to the results of the \( R_{\WR }=5.0\, R_{\sun } \)
models, and again the model with \( \dot{M}_{\WR }=0.6\times 10^{-5}\, \dot{M}_{\sun }\, \mathrm{yr}^{-1} \)
fits the observed \( I \), \( Q \), and \( U \) light curves at
the wavelength around \( 5630 \)~\AA~consistently. A summary of
the model parameters is given in Table~\ref{tab:ModelParameters}.


\begin{table*}

\caption{Model Summary \label{tab:ModelParameters}}

\vspace{0.3cm}
{\centering \begin{tabular}{lcccccc}
 &
&
&
&
&
&
\\
\hline
\hline 
MODEL&
 A&
B&
 C&
 D&
E&
 F\\
\hline
\( R_{\mathrm{WR}}\, \left[ R_{\sun }\right]  \)&
5.0&
5.0&
5.0&
2.5&
2.5&
2.5\\
\( L_{\mathrm{WR}}\, \left[ 10^{5\, }L_{\sun }\right]  \) &
2.0&
2.0&
2.0&
2.0&
2.0&
2.0\\
&
&
&
&
&
&
\\
\( \dot{M}_{\mathrm{WR}} \)~\( \left[ \times 10^{-5}\, M_{\sun }\, \mathrm{yr}^{-1}\right]  \)&
0.3&
0.6&
1.2&
0.3&
0.6&
1.2\\
\( V_{\infty } \)(WR)~\( \left[ \mathrm{km}\, \mathrm{s}^{-1}\right]  \)&
1700&
1700&
1700&
1700&
1700&
1700\\
\( f \)&
0.018&
0.05&
0.15&
0.025&
0.075&
0.225\\
&
&
&
&
&
&
\\
\( q\left( \lambda 5630\right)  \)&
0.52&
0.45&
0.40&
0.31&
0.31&
0.29\\
\( q\left( \lambda 22000\right)  \)&
1.15&
1.10&
1.05&
1.13&
0.96&
0.90\\
&
&
&
&
&
&
\\
\( R_{\mathrm{O}}\, \left[ R_{\sun }\right]  \)&
5.7&
7.2&
10.0&
5.5&
6.9&
8.6\\
&
&
&
&
&
&
\\
\( i\, \left[ \mathrm{deg}.\right]  \)&
80.5&
78.0&
73.0&
82.0&
79.5&
76.5\\
\hline
\end{tabular}\par}
\vspace{0.3cm}
\end{table*}


\subsubsection{Effects of the Bow Shock on Light Curves }

\label{subsubsec:BowshockEffect}

\begin{figure}
{\centering \resizebox*{1\columnwidth}{!}{\includegraphics{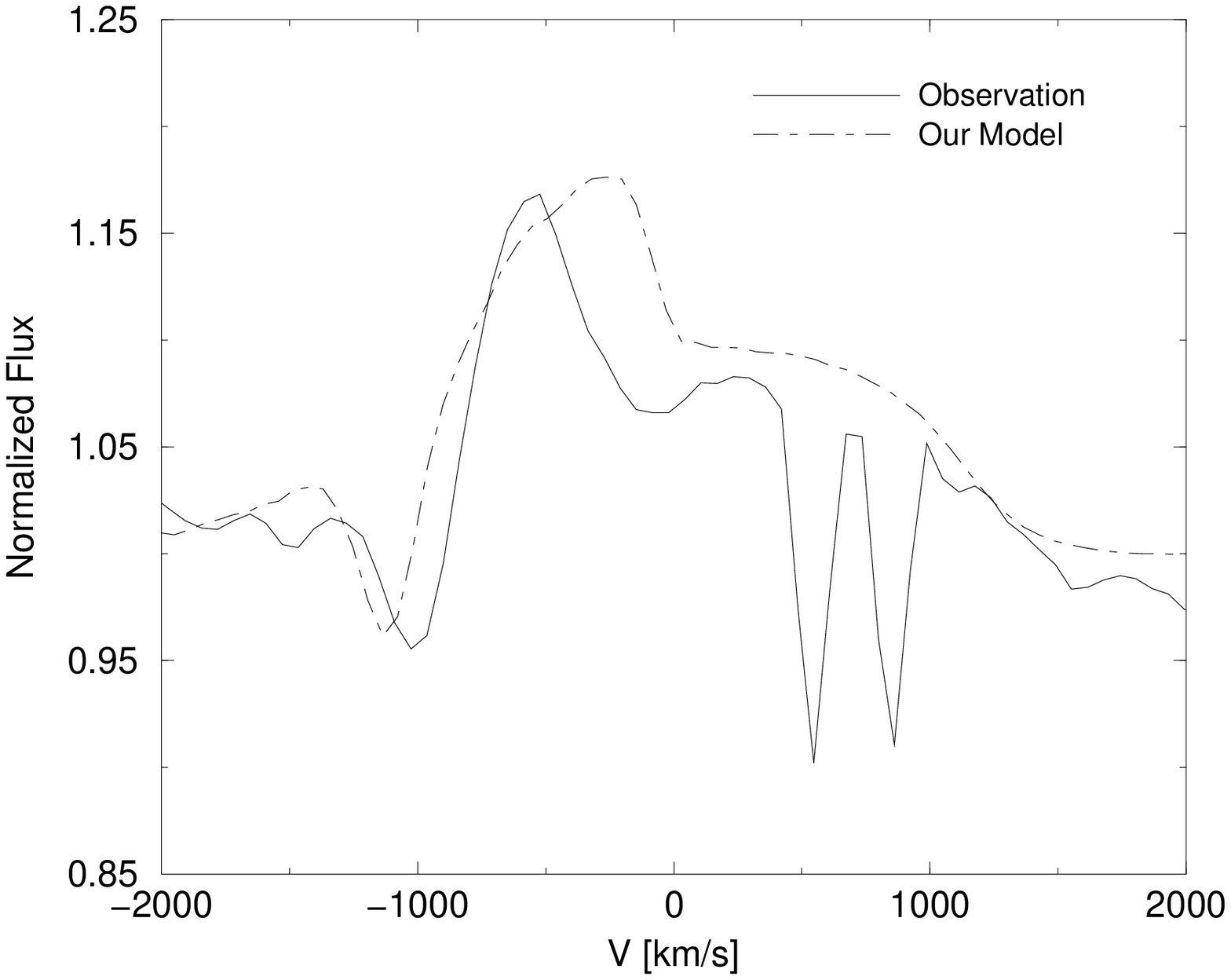}} \par}

\caption{\label{fig:CWfitHeIline} Spectral model of \HeI~\( 5876 \)~\AA~at
phase \( =0.419 \) where the emission from the bow shock is prominent. The
profile is fitted by adjusting \( \rho _{o} \) in Eq.~\ref{eq:CWdensity02}
to produce a similar amount of the excess emission as seen on the
top of the \HeI~\( 5876 \)~\AA~line. The tilt of the bow shock
\( \delta =15^{o} \) and open angle \( =90^{\arcdeg } \) \citep{marchenko:1994}
were used in the geometrical bow shock model introduced in \S~\ref{subsec:LineVarBowShockModel}.
To compute the profile, the emissivity and the opacity from Model
B were used in the 3-D Monte Carlo model. }
\end{figure}

We now consider the possible effects of the presence of the bow shock
region on our computed light curves (Figs.~\ref{fig:FinalFits} and
\ref{fig:FinalFits2}). To do so, the observed \HeI~\( 5876 \)~\AA~line
at the phase where the effect of the CW emission is most prominent
is modeled first. The spectrum taken at \( \phi =0.419 \) (see Fig.~\ref{fig:HeFits})
is chosen for this purpose. The profile at this phase is fitted by
adjusting \( \rho _{o} \) in Eq.~\ref{eq:CWdensity02} to produce
a similar amount of the excess emission as seen on the top of the
\HeI~\( 5876 \)~\AA~line. Since the true line opacity and emissivity
in the bow shock region are not known, this was done just to produce
the bow shock line emission seen in \HeI~\( 5876 \)~\AA. We do
not intend to perform a very rigorous fit of this line although that
would be a good topic for a future investigation. The result from
the 3-D Monte Carlo calculation for Model B is shown in Fig.~\ref{fig:CWfitHeIline}.
A tilt of the bow shock, \( \delta =15^{o} \), and open angle \( =90^{\arcdeg } \)
\citep{marchenko:1994} were used in this model. The fit shown in Fig.~\ref{fig:CWfitHeIline}
is reasonably good except for the flux level around \( V=200-500\, \mathrm{km}\, \mathrm{s}^{-1} \)
where the observed spectrum is affected by the O star absorption.
For simplicity, O star atmospheric absorption effects are not included
in this model; however, an appropriate \( q \) was used to produce
the correct O star continuum flux. The ratio the bow shock density
to the background density (\( \rho _{\mathrm{cw}}/\rho _{\mathrm{bg}} \)),
due to the W-R stellar wind, is found to be about 40 according to
this model. This ratio is similar to that of \citet{pittard:1999}.

\begin{figure}
{\centering \resizebox*{1\columnwidth}{!}{\includegraphics{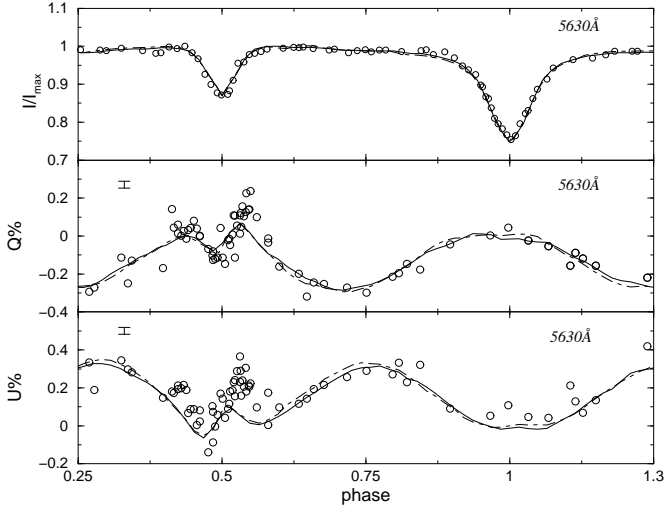}} \par}

\caption{\label{fig:LightCurveWithCW}Comparisons of the light curves
of Model B in Table~\ref{tab:ModelParameters} with and without the
bow shock. The opacity of the bow shock used here is obtained by fitting
\HeI~\( 5876 \)~\AA~at phase \( =0.419 \) as shown in Fig.~\ref{fig:CWfitHeIline}.
Solid line: Model B without the bow shock. Dot-Dash: Model with the
bow shock. The data points are indicated by circles. Within the range
of the error, the two light curve models are identical to each other.
The size of the typical error (arising from random fluctuations) for
the model \( Q \) and \( U \) polarization curves are shown on the
upper left hand corner of each plot. The size of the error for \( I \)
is too small to be shown on the plot.}
\end{figure}

With this bow shock model, the light curves (\( I \), \( Q \), and
\( U \)) at \( \lambda =5630 \)~\AA~ of Model B (\( \dot{M}_{\WR }=0.6\times 10^{-5}\, M_{\sun }\, \mathrm{yr}^{-1} \)
and \( R_{\WR }=5.0\, R_{\sun } \)) in Fig.~\ref{fig:FinalFits}
are recomputed. Fig.~\ref{fig:LightCurveWithCW} shows the result.
There is no significant effect of the bow shock region on the continuum
\( I \), \( Q \) and \( U \) light curves. The two models are identical
to each other within the range of the error. The same conclusion was
obtained for the other models shown in Figs.~\ref{fig:FinalFits}
and \ref{fig:FinalFits2}.

\subsubsection{Effect of the O-Star Wind using a Hydrodynamical Model}

\label{subsubsec:OstarWind}

To check the validity of ignoring the O star wind and the geometrical
bow shock model, the polarization light curves were computed using
the hydrodynamical model of \citet{pittard:1999}, which includes the
orbital motion of the stars and a parameterized form of radiative
driving. First, the opacity of the W-R star atmosphere was computed
with CMFGEN using the parameters adapted by \citet{pittard:1999}:
\( R_{\WR }=4.0\, R_{\sun } \), \( R_{\Ostar }=10.0\, R_{\sun } \),
\( L_{\WR }=1.40\times 10^{5}\, L_{\sun } \), \( L_{\Ostar }=2.4\times 10^{5}\, L_{\sun } \),
\( \dot{M}_{\WR }=9.72\times 10^{-6}\, M_{\sun }\, \mathrm{yr}^{-1} \),
\( \dot{M}_{\Ostar }=5.78\times 10^{-7}\, M_{\sun }\, \mathrm{yr}^{-1} \),
\( V^{\WR }_{\infty }=1800\, \mathrm{km}\, \mathrm{s}^{-1} \) and
\( V^{\Ostar }_{\infty }=2900\, \mathrm{km}\, \mathrm{s}^{-1} \).
The CMFGEN model with these parameters does not fit the observed helium
spectrum, but the opacity from this model will be used in the polarization
calculation to maintain the consistency with the hydrodynamical model.

\begin{figure}
{\centering \resizebox*{1\columnwidth}{!}{\includegraphics{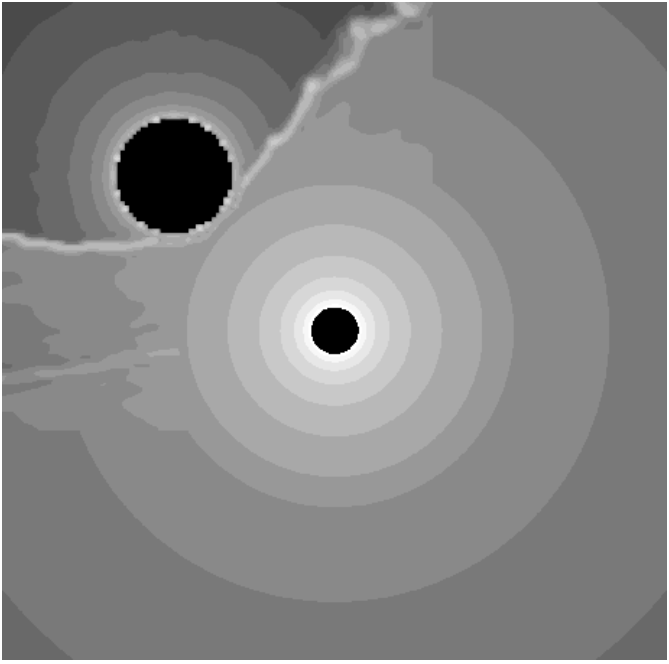}} \par}

\caption{\label{fig:stevens}This figure shows the density distribution
of the V444~Cyg model described in \S~\ref{subsubsec:OstarWind}.
The density, including the bow shock, near the O star (\emph{upper-left})
is from the hydrodynamical model of \citet{pittard:1999}, and is added
to the spherical density distribution of the W-R star (\emph{center})
derived from a CMFGEN model. The parameters used in this model are:
\( R_{\WR }=4\, R_{\sun } \), \( R_{\Ostar }=10\, R_{\sun } \) and
\( a=38\, R_{\sun } \). \( L_{\WR }=1.40\times 10^{5}\, L_{\sun } \),
\( L_{\Ostar }=2.4\times 10^{5}\, L_{\sun } \), \( \dot{M}_{\WR }=9.72\times 10^{-6}\, M_{\sun }\, \mathrm{yr}^{-1} \),
\( \dot{M}_{\Ostar }=5.78\times 10^{-7}\, M_{\sun }\, \mathrm{yr}^{-1} \),
\( V^{\WR }_{\infty }=1800\, \mathrm{km}\, \mathrm{s}^{-1} \) and
\( V^{\Ostar }_{\infty }=2900\, \mathrm{km}\, \mathrm{s}^{-1} \).}
\end{figure}

\begin{figure}
{\centering \resizebox*{1\columnwidth}{!}{\includegraphics{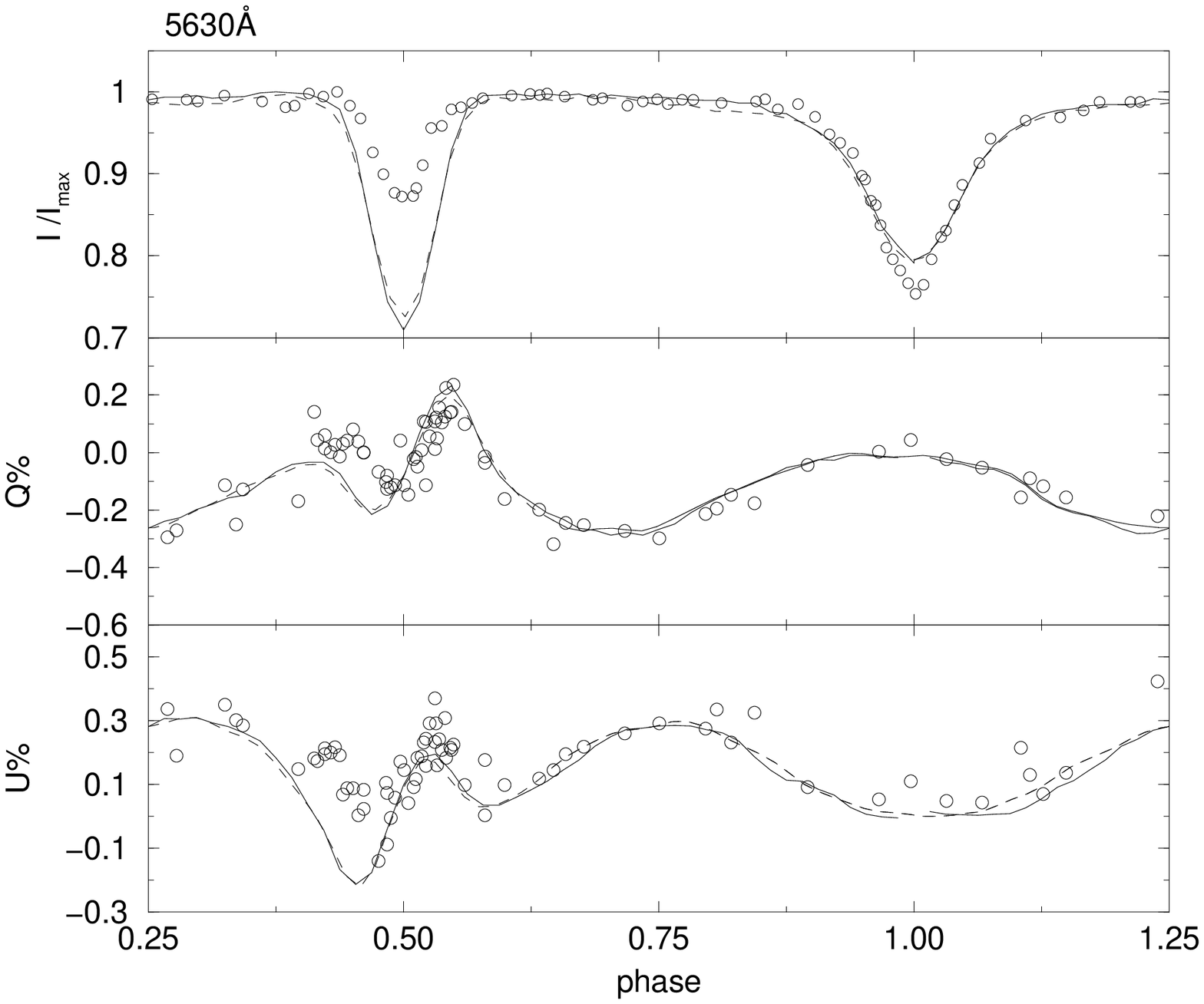}} \par}

\caption{\label{fig:StevensTest}The light curves are computed with
the parameters used in the hydrodynamical calculation of \citet{pittard:1999}.
They are: \( R_{\WR }=4.0\, R_{\sun } \), \( R_{\Ostar }=10.0\, R_{\sun } \),
\( L_{\WR }=1.40\times 10^{5}\, L_{\sun } \), \( L_{\Ostar }=2.4\times 10^{5}\, L_{\sun } \),
\( \dot{M}_{\WR }=9.72\times 10^{-6}\, M_{\sun }\, \mathrm{yr}^{-1} \)
and \( \dot{M}_{\Ostar }=5.78\times 10^{-7}\, M_{\sun }\, \mathrm{yr}^{-1} \).
The solid lines are from the model without the bow shock and the O
star wind. The dashed lines are the model with the O star wind and
the bow shock from the hydrodynamical model of \citet{pittard:1999}
(Fig.~\ref{fig:stevens}). No significant differences between the
two models are seen in this figure.}
\end{figure}

Using this W-R star atmosphere model, the following two models are
computed: 1.\,a model without the bow shock and the O star wind, 2.\,a
model with the O star wind and the bow shock from the hydrodynamical
calculation of \citet{pittard:1999}. The density distribution for
the second case is shown in Fig.~\ref{fig:stevens}. To assign the
opacity and emissivity in the O star wind and the bow shock regions,
we assumed that the thermal opacity and the emissivity are proportional
to the square of the density, and the electron scattering opacity
is proportional to the density. Then the opacity and the emissivity
are scaled with those from the W-R star atmosphere model. The light
curves computed for these models are shown in Fig.~\ref{fig:StevensTest}.
No significant differences between the two models are seen in this
figure. This validates one of our main model assumptions, i.e., the
O star wind does not significantly affect the continuum polarization
curves. It also assures that the details of the bow shock structure
are not important for predicting the \emph{continuum} polarization
curves though they are very important for a line calculation. 

The model light curves in Fig.~\ref{fig:StevensTest} do not fit
the observed light curves. By increasing the inclination angle, the
depth of the eclipses (at \( \phi =0 \) and \( 0.5 \)) will increase;
however, without adjusting other parameters, it is impossible to decrease
the depth of the secondary eclipse. It is clearly seen from the location
of minima in the eclipsing \( Q \) and \( U \) light curves (\( \phi \approx 0.46 \))
that the O star radius is too large. Surprisingly, the amplitudes
of the \( Q \) and \( U \) double sine curves fit the observation
well with \( \dot{M}_{\WR }=9.72\times 10^{-6}\, M_{\sun }\, \mathrm{yr}^{-1} \).
This seems contradictory to our earlier analysis, but is due to the
inconsistent W-R model, which does not fit the observed helium spectrum,
being used in the polarization calculations. Since the parameters
from the model of \citet{pittard:1999} are fixed, \( q=L_{\WR }/L_{\Ostar }=0.583 \)
is also fixed in these models. If \( q=0.38 \) is used by increasing
\( L_{\Ostar } \) instead, the spectral model will fit the observed
helium spectrum, but the higher \( L_{\Ostar } \) results in larger
amplitudes of the double-sine polarization curves which then no longer
fit the observation.

In summary, from this analysis, we conclude that the mass-loss rate
of the W-R star in V444~Cyg is \( 0.6\left( \pm 0.2\right) \times 10^{-5}\, M_{\sun }\, \mathrm{yr}^{-1} \).
The size of the error in \( \dot{M} \) is estimated from the light
curves in Figs.~\ref{fig:FinalFits} and \ref{fig:FinalFits2}. The
mass-loss rate determined here is consistent with the values of STL2, and
it is slightly higher than that of \citet{underhill:1990}. (See Table~\ref{tab:V444ParaPublished}
for other published values.) The range of the volume filling factor
for the W-R star atmosphere is estimated to be \( 0.050-0.075 \)
with the corresponding range of the W-R star radius, \( 5.0-2.5\, R_{\sun } \).
We also found that the presence of the bow shock, the exact details
of the bow shock, and the presence of the O star wind are unimportant
for the continuum fits, and they do not affect the mass-loss rate
estimate given here.

\section{Discussion}

\label{sec:Discussion}

\subsection{Infrared Light Curves}

\begin{figure}
{\centering \resizebox*{1\columnwidth}{!}{\includegraphics{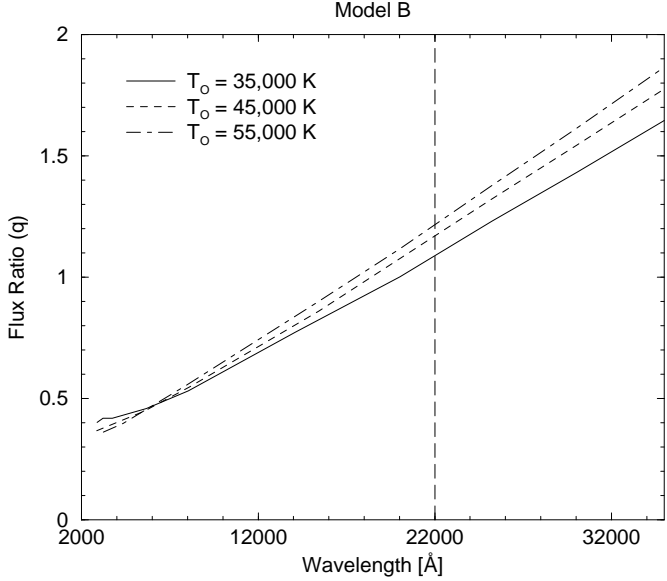}} \par}

\caption{\label{fig:qRatiovsLambda}The continuum flux ratio \( \left( q=F_{\WR }/F_{\Ostar }\right)  \)
of Model B as a function of wavelength for three different O star
effective temperatures. The lines were normalized at \( \lambda =5630 \)\AA~
where the luminosity ratio \( q=0.45 \) was determined from the spectroscopic
model fit (Fig.~\ref{fig:5RsBestFits}). }
\end{figure}

In order to check the consistency of our best fit models (Models B
and E with \( \dot{M}_{\WR }=0.6\times 10^{-5}\, M_{\sun }\mathrm{yr}^{-1} \))
in \S~\ref{subsec:MassLoss} with some other observational aspects,
the infrared light curves (at \( \lambda =2.2\, \mu \mathrm{m} \))
were also computed. In order to calculate the IR light curves, the
monochromatic luminosity ratio \( q=L_{\WR }/L_{\Ostar } \) at \( \lambda =2.2\, \mu \mathrm{m} \)
must first be estimated. In Model B, \( q \) was determined to be
\( 0.45 \) at \( \lambda =5630 \)~\AA~ in the previous section.
By normalizing the O star flux with this value of \( q \) at \( \lambda =5630 \)~\AA,
\( q \) for the wavelengths between \( 3000 \)~\AA~and \( 33,000 \)~\AA~is
plotted in Fig.~\ref{fig:qRatiovsLambda} for three different O star
effective temperatures. The figure shows that the O star is brighter
than the W-R star at \( \lambda =5630 \)~\AA, but the W-R star is
brighter than the O star after \( \lambda \sim 20,000 \)~\AA. In
model B, an O star temperature of \( T_{\Ostar }=35,000\, \mathrm{K} \)
was used; hence, the corresponding value of \( q \) at \( \lambda =2.2\, \mu m \)
is 1.10 according to the graph. For models with higher O star temperatures,
the values of \( q \) are slightly higher.

The resulting model IR light curves are shown in the right hand side
of Figs.~\ref{fig:FinalFits} and \ref{fig:FinalFits2} along with
Q and U polarization light curves for \( \dot{M}_{\WR }=0.3\times 10^{-5}\, M_{\sun } \),
\( \dot{M}_{\WR }=0.6\times 10^{-5}\, M_{\sun }\, \mathrm{yr}^{-1} \)
and \( \dot{M}_{\WR }=1.2\times 10^{-5}\, M_{\sun }\, \mathrm{yr}^{-1} \).
The figure also includes the observed IR light curve from the broadband
(\( K \)) photometry obtained by \citet{hartmann:1978}. The depth
of the primary eclipse (at \( \phi =0 \) or \( 1 \) when the W-R
star is in front of the O star) is fitted well for both \( R_{\WR }=2.5\, R_{\sun } \)
and \( 5.0\, R_{\sun } \) models, but the secondary eclipse modeled
(at \( \phi =0.5 \) when the O star is in front of the W-R star)
is shallower than the observed IR light curve. We also notice that
the maximum \( I \) flux level (\( I_{\mathrm{max}} \)) in the observed
light curve is not well defined. \citet{cherepashchuk:1984} and \citet{hamann:1992}
had a similar problem -- the primary and secondary eclipses are too
shallow in their light curve model. While the light curve model of
\citet{hamann:1992} did not fit the depth of both eclipses, our model
is consistent with at least the primary eclipse. 

\citet{cherepashchuk:1984} discussed two possible causes for this
problem: a).~the existence of {}``third light'' and b).~the result
of ignoring the reflection effect. They noted that one possible source
of third light is the contribution of emission lines in the broad-band
photometry at \( 2.2\, \mu \mathrm{m} \) as originally mentioned
by \citet{hartmann:1978}, but we now expect this effect to be generally
small (unless the \HeI~\( 2.06\, \mu \mathrm{m} \) emission line
is abnormally large). To examine this possibility, we have run a model
which includes the reflection by the disk of the O star. However we
find that it does not affect the infrared and optical light curves.
\citet{cherepashchuk:1984} also conjectured that the increase in the
depth and the width of the secondary eclipse (from optical to IR)
is a result of clumps in the WR stellar wind.

If the wrong flux ratio at \( \lambda =2.2\, \mu \mathrm{m} \) is
used in our model, the secondary eclipse could become shallower at
this wavelength. The differences in the depth of the secondary eclipse
in Figs.~\ref{fig:FinalFits} and \ref{fig:FinalFits2} are approximately
\( 8 \)\% and \( 5 \)\%. This could be explained by the wrong O
star temperature used in the model. According to Fig.~\ref{fig:qRatiovsLambda},
\( q \) at \( 2.2\, \mu \mathrm{m} \) also increases about \( 8 \)\%
if we use \( T_{\Ostar }=45,000\, \mathrm{K} \) although this would
not fit the absorption components of the observed spectrum.

\begin{table}

\caption{Summary of the Best Fit Models \label{tab:BestFitModels}}

\vspace{0.3cm}
{\centering \begin{tabular}{lcc}
 &
&
\\
\hline
\hline 
&
Model B&
Model E\\
\hline
\( R_{\mathrm{WR}}\, \left[ R_{\sun }\right]  \)&
5.0&
2.5\\
\( R_{\mathrm{O}}\, \left[ R_{\sun }\right]  \)&
7.2&
6.9\\
&
&
\\
\( \dot{M}_{\mathrm{WR}} \) \( \left[ \times 10^{-5}\, M_{\sun }\, \mathrm{yr}^{-1}\right]  \)&
0.6&
0.6\\
\( f \) &
0.05&
0.075\\
&
&
\\
\( q\left( \lambda 5630\right)  \)&
0.45&
0.31\\
\( q\left( \lambda 22000\right)  \)&
1.10&
0.96\\
&
&
\\
\( i\, \left[ \mathrm{deg}.\right]  \)&
78.0&
79.5\\
&
&
\\
\( M_{v} \) (O star)&
-5.20&
-4.85\\
\( M_{v} \) (W-R star)&
-4.33&
-3.58\\
\( M_{v} \) (O+W-R)&
-5.60&
-5.15\\
\( m_{v} \) (O+W-R)\( \left( *\right)  \)&
8.38&
8.84\\
\( DM \) \( \left( **\right)  \)&
11.1&
10.6\\
\hline
\end{tabular}\par}
\vspace{0.3cm}

\( \left( *\right)  \) Assuming a distance of 1.7 kpc and \( E_{b-v}=0.68 \)
\citet{lundstrom:1984} -- They assumed the intrinsic color to be \( \left( b-v\right) _{o}=-0.30 \).

\( \left( **\right)  \) Using the observed value of \( m_{v}=8.27 \)
and \( A_{v}=2.79 \) from \cite{lundstrom:1984}.

\end{table}

\subsection{Polarization Curves near the Secondary Eclipse}

In both Model B (Fig.~\ref{fig:FinalFits}) and Model E (Fig.~\ref{fig:FinalFits2}),
the amplitude of the \( U \) polarization variation near the secondary
eclipse (\( \phi =0.5 \)) underestimates the observed level. A larger
polarization amplitude can be obtained if the radius of the O star
is increased, although this will result in an inconsistent depth of
the secondary eclipse. A possible cause of this inconsistency is the
inaccurate stellar wind structure very close to the core of the W-R
star. For all the models in Table~\ref{tab:ModelParameters}, \( \beta _{1}=1 \)
was used in the modified beta-velocity law \citep{hillier:1999} (see
Eq.~\ref{eq:modefiedBetaVelocity}) where \( \beta _{1} \) is equivalent
to \( \beta  \) in the classic beta-velocity law for the {}``inner''
part of the stellar wind. To see if changing the structure of the
inner W-R stellar wind affects the computed continuum polarization
level, a model with \( \beta _{1}=2 \) (a slower wind) is also calculated.
Unfortunately, the amplitude of the polarization curves (not shown
here) near \( \phi =0.5 \) did not change significantly for the new
\( \beta _{1} \) value. Changing the value of \( \beta _{1} \) also
did not influence the depth of the secondary eclipse in the IR light
curve. 

We note that STL2 were able to fit the observed \( Q \) and \( U \)
variations around \( \phi =0.5 \). There are two reasons for their
success: 1.~they did not fit the observed \( I \) light curve consistently,
and 2.~they use a fixed orbital inclination angle. In their fitting
procedure, the radius of the O star was used as a fitting parameter,
but changing the O star radius changes the depth of the eclipse around
\( \phi =0.5 \). Consequently, the inclination angle must also be
adjusted when the O star radius is changed, to maintain a consistency
with the observed \( I \) light curve. We should also note that the
amplitudes of polarization curves around \( \phi =0.5 \) are very
sensitive to the orbital inclination. STL2 adopted \( i=78.9^{\arcdeg } \)
from \citet{robert:1990}. However, if instead \( i=78^{\arcdeg } \)
from \citet{cherepashchuk:1975} were used, they would obtain a fairly
different set of parameters from the ones derived in their paper.
Fortunately, the amplitudes of the double sine curves are not very
sensitive to the adopted inclination angle. Considering the relative
sensitivity of different methods on the inclination angle, the mass-loss
rate estimation by fitting the double sine curves is more reliable
than the one obtained by fitting the polarization variations near
\( \phi =0.5 \).

\subsection{Magnitudes and Distance Moduli}

For each model in Figs.~\ref{fig:FinalFits} and \ref{fig:FinalFits2},
the absolute magnitude (\( M_{v} \)) of the system (W-R + O) is calculated,
and placed in Table~\ref{tab:BestFitModels}. \( M_{v}=-5.60 \) from
Model B is very similar to \( M_{v}=-5.7 \) of \citet{lundstrom:1984},
but is far from the value of \citet{hamann:1992}, \( M_{v}=-4.2 \).
Model E gives \( M_{v}=-5.15 \) which is slightly different from
the values estimated by \citet{lundstrom:1984} and \citet{hamann:1992}.
Using the observed \( m_{v}=8.27 \) and \( A_{v}=2.79 \) from \citet{lundstrom:1984},
we have also estimated the distance modulus (\( DM \)) for each model,
and summarized them in the same table. The values of \( DM \) from
Model B and Model E are \( 11.1 \) and \( 10.6 \) respectively.
The former (Model B) agrees with the value given by \citet{lundstrom:1984}
(\( DM=11.01\pm 0.20 \)), but it is slightly smaller than the value
estimated by \citet{marchenko:1997} (\( DM=11.27-11.47 \)). 

The absolute magnitudes of the O star corresponding to Models B and
E are also listed in Table~\ref{tab:BestFitModels}. They are \( M_{v}=-5.20 \)
and \( M_{v}=-4.85 \) respectively. According to \citet{howarth:1989},
the mean absolute magnitude of O6 V stars is \( M_{v}=-5.2 \), and
that for O6 III is \( M_{v}=-6.6 \). Our magnitude of Model B agrees
with the mean of their O6 V stars. Similarly, the average absolute
magnitude of observed WN5 stars is \( M_{v}=-4.1\left( \pm 0.8\right)  \)
\citep{vanderhucht:2001} which also agrees with our values \( M_{v}=-4.33 \)
(Model B) and \( M_{v}=-3.58 \) (Model E). 

The derived O star radii are \( R_{\Ostar }=7.2\, R_{\sun } \)(Model
B) and \( R_{\Ostar }=6.9\, R_{\sun } \)(Model E). Both values are
slightly smaller than the recent value \( R_{O}=8.5\pm 1.0\, R_{\sun } \)
estimated by STL2, and are not consistent with \( R_{\Ostar }=10.0\, R_{\sun } \)
\citep{cherepashchuk:1984} and \( R_{\Ostar }=8.4-9.3\, R_{\sun } \) \citep{moffat:1996}.
The rather small O star radii of the models and the model \( M_{v} \)
values compared to the mean observed \( M_{v} \) suggest that the
O star is more likely to be a main-sequence star rather than a giant
star. A similar conclusion about the O star luminosity class was reached
by STL2. 

\citet{vanbeveren:1998} quoted \citet{cherepashchuk:1975}, \char`\"{}The
observed mass of the WNE component is \( 9\, M_{\sun } \); its progenitor
must therefore have had a mass of \( \sim 30\, M_{\sun } \). Thus,
the age of the binary is about 7 million years. The OB companion is
an O6 star. Since the age of a normal O6 star is about 1-2 million
years, the OB star must have been rejuvenated, i.e. mass transfer
must have occurred.\char`\"{} This argument suggests that \emph{the
O star is not a normal O star}. According to \citet{vanbeveren:1998},
a quasi-conservative Roche lobe overflow model can produce a system
which resembles V444~Cyg. At the end of the mass transfer, when the
secondary has been entirely mixed, the O6 component may be an \emph{over-luminous}
star with \( M_{\Ostar }=26\, M_{\sun } \). Contrary to the last
remark, our best fit models (Models B and E) suggest that the absolute
magnitude of the O star is similar to the average of O6 V stars, as
discussed earlier. 

Unfortunately, we can not strongly conclude that either Model B or
E is a more favorable model based on the comparisons with the observed
magnitudes, distance moduli and the results from earlier works.

\section{Conclusions}

\label{sec:Conclusion}

Using the 3-D Monte Carlo model developed in Paper I, combined with
the multi-line non-LTE radiative model of \citet{hillier:1998}, we
first investigated the basic behavior of the WN star diagnostic lines
(\HeII~\( 5412 \)~\AA~and \HeI~\( 5876 \)~\AA) as a function
of the phase of V444~Cyg (Figs.~\ref{fig:CWemission}--\ref{fig:HeFits}).
Since the emission lines from the W-R star envelope are contaminated
by the O star atmospheric absorption lines and the line emission from
the shock-heated region, we first examined at which phase the profile
shape of helium lines, especially \HeI~\( 5876 \)~\AA, is least
affected by the contaminations. We then fit the \HeI~\( 5876 \)~\AA~profile
with the multi-line non-LTE radiative transfer model with spherical
geometry to obtain a realistic model for the W-R star. The best orbital
phase for this purpose is found to be around \( \phi =0.8 \).

A summary of the derived parameters of V444~Cyg with our best fit
models (Models B \& E) is given in Table~\ref{tab:BestFitModels}.
Both models show excellent agreement with the observed light curves
of \( I \), \( Q \) and \( U \) Stokes parameters (Figs.~\ref{fig:FinalFits}
and \ref{fig:FinalFits2}). They also fit the spectrum of \HeII~\( 5412 \)~\AA~and
\HeI~\( 5876 \)~\AA~very well (Figs.~\ref{fig:5RsBestFits} and
\ref{fig:2halfRsBestFits}). The effect of ignoring the presence of
the bow shock in the continuum polarization calculation was also considered.
This was done by first approximating the opacity in the bow shock
region via fitting \HeI~\( 5876 \)~\AA~line at \( \phi =0.419 \)
where the emission from the bow shock is most prominent (Fig~\ref{fig:CWfitHeIline}).
According to this model, the bow shock region has little or no effect
on the continuum light curves at \( \lambda =5630 \)\AA~for all
\( I \), \( Q \) and \( U \) (Fig.~\ref{fig:LightCurveWithCW}).
Similarly, the presence of the O-star wind and the exact details of
the bow shock are found to be unimportant for the continuum fits (Fig.~\ref{fig:StevensTest}).
We also found that the spectroscopic solution, when combined with
the strong O star spectrum, is very insensitive to the value of the
W-R core radius (c.f., Fig.~\ref{fig:5RsBestFits} and Fig.~\ref{fig:2halfRsBestFits}).

Small discrepancies are seen in the oscillation of the \( Q \) and
\( U \) light curves (Figs.~\ref{fig:FinalFits} and \ref{fig:FinalFits2})
near the secondary eclipse (\( \phi =0.5 \)), which are possibly
due to the lack of knowledge of the exact opacity and emissivity in
the bow shock region. This discrepancy could be a result of
the wrong combination of the orbital inclination angle and the O star
radius in our model. The shape of the polarization curves near \( \phi =0.5 \)
is very sensitive to the inclination angle, and is difficult to fit
consistently with other observational features.

The most noticeable difference between our best model and the observations
is in the IR light curves (\( 2.2\, \mu \mathrm{m} \)) in Figs.~\ref{fig:FinalFits}
and \ref{fig:FinalFits2}. The model correctly predicts the depth
of the primary eclipse, but that of the secondary eclipse is too shallow.
We argued that the difference could be caused by the use of the wrong
monochromatic luminosity ratio (\( q \)) at \( \lambda =2.2\, \mu \mathrm{m} \). 

The mass-loss rate of the W-R component determined in our analysis
is \( \dot{M}_{\WR }=0.6\left( \pm 0.2\right) \times 10^{-5}\, M_{\sun }\, \mathrm{yr}^{-1} \),
and this conclusion is insensitive to the model radius of the W-R
star. The mass-loss rate of the W-R star obtained by STL2, \( \dot{M}_{\WR }=0.6\times 10^{-5}\, M_{\sun }\, \mathrm{yr}^{-1} \),
is in good agreement with our value derived here. Our derived mass-loss
rate lies between the values obtained by the orbital period change
method (Table~\ref{tab:V444ParaPublished}). The fits did not allow
a unique value for the radius of the W-R star to be derived. The range
of the volume filling factor for the W-R star atmosphere is estimated
to be \( 0.050-0.075 \) for the corresponding range of the W-R star
radius, \( 5.0-2.5\, R_{\sun } \). 

Future work includes: 1.~obtaining improved IR light curves for both
models and observations, 2.~fitting and modeling the UV spectrum
and polarization, 3.~checking the consistency between the polarization
model and the hydrodynamical model more carefully, 4.~investigating
the cause for the asymmetry of the polarization light curve around
the secondary eclipse, and 5.~obtaining a tighter constraint on \( q \).
Future interferometer observations of this system will also prove
invaluable to our further understanding.

\begin{acknowledgements}
We would like to thank Dr.~S.~Marchenko and Dr.~A.~F.~J\@.~Moffat
for providing us the helium spectrum sequence data and for giving
us very constructive comments on the manuscript. This research was
supported by NASA grant NAGW-3828.
\end{acknowledgements}

%

%

%



\begin{thebibliography}{72}
\expandafter\ifx\csname natexlab\endcsname\relax\def\natexlab#1{#1}\fi

\bibitem[{Abbott {et~al.}(1981)Abbott, Bieging, \& Churchwell}]{abbott:1981}
Abbott, D.~C., Bieging, J.~H., \& Churchwell, E. 1981, \apj, 250, 645

\bibitem[{{Antokhin} {et~al.}(1995){Antokhin}, {Marchenko}, \&
  {Moffat}}]{antokhin:1995}
{Antokhin}, I.~I., {Marchenko}, S.~V., \& {Moffat}, A.~F.~J. 1995, in IAU
  Symp., Vol. 163, {Wolf-Rayet Stars: Binaries; Colliding Winds; Evolution},
  ed. K.~A. van~der Hucht \& P.~M. Williams (Dordrecht: Kluwer), 520

\bibitem[{{Bartzakos} {et~al.}(2001){Bartzakos}, {Moffat}, \&
  {Niemela}}]{bartzakos:2001}
{Bartzakos}, P., {Moffat}, A.~F.~J., \& {Niemela}, V.~S. 2001, \mnras, 324, 33

\bibitem[{Beals(1944)}]{beals:1944}
Beals, C. 1944, \mnras, 104, 205

\bibitem[{{Blondin} \& {Marks}(1996)}]{blondin:1996}
{Blondin}, J.~M. \& {Marks}, B.~S. 1996, New Astronomy, 1, 235

\bibitem[{Brown {et~al.}(1978)Brown, McLean, \& Emslie}]{brown:1978}
Brown, J.~C., McLean, I.~S., \& Emslie, A.~G. 1978, \aap, 68, 415

\bibitem[{{Canto} {et~al.}(1996){Canto}, {Raga}, \& {Wilkin}}]{canto:1996}
{Canto}, J., {Raga}, A.~C., \& {Wilkin}, F.~P. 1996, \apj, 469, 729

\bibitem[{{Castor} {et~al.}(1975){Castor}, {Abbott}, \& {Klein}}]{castor:1975}
{Castor}, J.~I., {Abbott}, D.~C., \& {Klein}, R.~I. 1975, \apj, 195, 157

\bibitem[{{Chandrasekhar}(1961)}]{chandrasekhar:1961}
{Chandrasekhar}, S. 1961, {Hydrodynamic and Hydromagnetic Stability} (Oxford:
  Oxford Univ.~Press)

\bibitem[{Cherepashchuk(1975)}]{cherepashchuk:1975}
Cherepashchuk, A.~M. 1975, Soviet Astron., 19, 727

\bibitem[{Cherepashchuk {et~al.}(1984)Cherepashchuk, Eaton, \&
  Khaliullin}]{cherepashchuk:1984}
Cherepashchuk, A.~M., Eaton, J.~A., \& Khaliullin, K.~F. 1984, \apj, 281, 774

\bibitem[{Cherepashchuk {et~al.}(1995)Cherepashchuk, Koenigsberger, Marchenko,
  \& Moffat}]{cherepashchuk:1995}
Cherepashchuk, A.~M., Koenigsberger, G., Marchenko, S.~V., \& Moffat, A.~F.~J.
  1995, \aap, 293, 142

\bibitem[{{Corcoran} {et~al.}(1996){Corcoran}, {Stevens}, {Pollock}, {Swank},
  {Shore}, \& {Rawley}}]{corcoran:1996}
{Corcoran}, M.~F., {Stevens}, I.~R., {Pollock}, A. M.~T., {Swank}, J.~H.,
  {Shore}, S.~N., \& {Rawley}, G.~L. 1996, \apj, 464, 434

\bibitem[{Forbes {et~al.}(1992)Forbes, English, de~Robertis, \&
  Dawson}]{forbes:1992}
Forbes, D., English, D., de~Robertis, M.~M., \& Dawson, P.~C. 1992, \aj, 103,
  916

\bibitem[{{Gayley} {et~al.}(1997){Gayley}, {Owocki}, \&
  {Cranmer}}]{gayley:1997}
{Gayley}, K.~G., {Owocki}, S.~P., \& {Cranmer}, S.~R. 1997, \apj, 475, 786

\bibitem[{{Girard} \& {Willson}(1987)}]{girard:1987}
{Girard}, T. \& {Willson}, L.~A. 1987, \aap, 183, 247

\bibitem[{Hamann {et~al.}(1995)Hamann, Koesterke, \& Wessolowski}]{hamann:1995}
Hamann, W.-R., Koesterke, L., \& Wessolowski, U. 1995, \aap, 299, 151

\bibitem[{{Hamann} \& {Schwarz}(1992)}]{hamann:1992}
{Hamann}, W.-R. \& {Schwarz}, E. 1992, \aap, 261, 523

\bibitem[{Hartmann(1978)}]{hartmann:1978}
Hartmann, L. 1978, \apj, 221, 193

\bibitem[{{Herald} {et~al.}(2001){Herald}, {Hillier}, \&
  {Schulte-Ladbeck}}]{herald:2001}
{Herald}, J.~E., {Hillier}, D.~J., \& {Schulte-Ladbeck}, R.~E. 2001, \apj, 548,
  932

\bibitem[{Hillier(1989)}]{hillier:1989}
Hillier, D.~J. 1989, \apj, 347, 392

\bibitem[{Hillier \& Miller(1998)}]{hillier:1998}
Hillier, D.~J. \& Miller, D.~L. 1998, \apj, 496, 407

\bibitem[{Hillier \& Miller(1999)}]{hillier:1999}
---. 1999, \apj, 519, 354

\bibitem[{Howarth \& Prinja(1989)}]{howarth:1989}
Howarth, I.~D. \& Prinja, R.~K. 1989, \apjs, 69, 527

\bibitem[{Howarth \& Schmutz(1992)}]{howarth:1992}
Howarth, I.~D. \& Schmutz, W. 1992, \aap, 261, 503

\bibitem[{{Huang} \& {Weigert}(1982)}]{huang:1982}
{Huang}, R.~Q. \& {Weigert}, A. 1982, \aap, 112, 281

\bibitem[{Khaliullin {et~al.}(1984)Khaliullin, Khaliullina, \&
  Cherepashchuk}]{khaliullin:1984}
Khaliullin, K.~F., Khaliullina, A.~I., \& Cherepashchuk, A.~M. 1984, Soviet
  Astron.~Lett., 10, 250

\bibitem[{{Kron} \& {Gordon}(1943)}]{kron:1943}
{Kron}, G.~E. \& {Gordon}, K.~C. 1943, \apj, 97, 311

\bibitem[{Kuhi(1968)}]{kuhi:1968}
Kuhi, L.~V. 1968, \apj, 152, 89

\bibitem[{{Kurosawa} \& {Hillier}(2001)}]{kurosawa:2001a}
{Kurosawa}, R. \& {Hillier}, D.~J. 2001, \aap, 379, 336

\bibitem[{{Lamontagne} {et~al.}(1996){Lamontagne}, {Moffat}, {Drissen},
  {Robert}, \& {Matthews}}]{lamontagne:1996}
{Lamontagne}, R., {Moffat}, A.~F.~J., {Drissen}, L., {Robert}, C., \&
  {Matthews}, J.~M. 1996, \aj, 112, 2227

\bibitem[{{Leitherer}(1988)}]{leitherer:1988}
{Leitherer}, C. 1988, \apj, 326, 356

\bibitem[{{L\"{u}hrs}(1997)}]{luhrs:1997}
{L\"{u}hrs}, S. 1997, \pasp, 109, 504

\bibitem[{Lundstr\"{o}m \& Stenholm(1984)}]{lundstrom:1984}
Lundstr\"{o}m, I. \& Stenholm, B. 1984, \aap, 58, 163

\bibitem[{Luo {et~al.}(1990)Luo, McCray, \& Mac~Low}]{luo:1990}
Luo, D., McCray, R., \& Mac~Low, M. 1990, \apj, 362, 267

\bibitem[{{Maeda} {et~al.}(1999){Maeda}, {Koyama}, {Yokogawa}, \&
  {Skinner}}]{maeda:1999}
{Maeda}, Y., {Koyama}, K., {Yokogawa}, J., \& {Skinner}, S. 1999, \apj, 510,
  967

\bibitem[{Marchenko {et~al.}(1997)Marchenko, Moffat, Eenens, Cardona,
  Echevarria, \& Hervieux}]{marchenko:1997}
Marchenko, S.~V., Moffat, A.~F.~J., Eenens, P.~R.~J., Cardona, O., Echevarria,
  J., \& Hervieux, Y. 1997, \apj, 422, 810

\bibitem[{Marchenko {et~al.}(1994)Marchenko, Moffat, \&
  Koenigsberger}]{marchenko:1994}
Marchenko, S.~V., Moffat, A.~F.~J., \& Koenigsberger, G. 1994, \apj, 422, 810

\bibitem[{Meynet {et~al.}(1994)Meynet, Maeder, Schaller, Schaerer, \&
  Charbonnel}]{meynet:1994}
Meynet, G., Maeder, A., Schaller, G., Schaerer, D., \& Charbonnel, C. 1994,
  \aaps, 103, 97

\bibitem[{Moffat {et~al.}(1982)Moffat, Firmani, McLean, \&
  Seggewiss}]{moffat:1982}
Moffat, A.~F.~J., Firmani, C., McLean, I.~S., \& Seggewiss, W. 1982, in IAU
  Symp., Vol.~99, Wolf-Rayet Stars: Observations, Physics, Evolution, ed.
  C.~W.~H. de~Loore \& A.~J. Willis. (Dordrecht: Kluwer), 577

\bibitem[{{Moffat} \& {Marchenko}(1996)}]{moffat:1996}
{Moffat}, A. F.~J. \& {Marchenko}, S.~V. 1996, \aap, 305, L29

\bibitem[{M\"{u}nch(1950)}]{munch:1950}
M\"{u}nch, G. 1950, \apj, 112, 266

\bibitem[{{Myasnikov} \& {Zhekov}(1993)}]{myasnikov:1993}
{Myasnikov}, A.~V. \& {Zhekov}, S.~A. 1993, \mnras, 260, 221

\bibitem[{{Najarro} {et~al.}(1997){Najarro}, {Hillier}, \&
  {Stahl}}]{najarro:1997}
{Najarro}, F., {Hillier}, D.~J., \& {Stahl}, O. 1997, \aap, 326, 1117

\bibitem[{Nugis {et~al.}(1998)Nugis, Crowther, \& Willis}]{nugis:1998}
Nugis, T., Crowther, P.~A., \& Willis, A.~J. 1998, \aap, 333, 956

\bibitem[{Piirola \& Linnaluoto(1988)}]{piirola:1988}
Piirola, V. \& Linnaluoto, S. 1988, in Polarized Radiation of Circumstellar
  Origin, ed. G.~V. Coyne, A.~M. {Magalh\~{a}es}, A.~F.~J. Moffat, R.~E.
  Schulte-Ladbeck, S.~Tapia, \& D.~T. Wickramasinghe (Vatican City State:
  Vatican Observatory), 655

\bibitem[{Pittard(1998)}]{pittard:1998}
Pittard, J.~M. 1998, \mnras, 300, 479

\bibitem[{{Pittard} \& {Corcoran}(2002)}]{pittard:2002}
{Pittard}, J.~M. \& {Corcoran}, M.~F. 2002, \aap, 383, 636

\bibitem[{{Pittard} \& {Stevens}(1997)}]{pittard:1997}
{Pittard}, J.~M. \& {Stevens}, I.~R. 1997, \mnras, 292, 298

\bibitem[{Pittard \& Stevens(1999)}]{pittard:1999}
Pittard, J.~M. \& Stevens, I.~R. 1999, in IAU Symp., Vol. 193, Wolf-Rayet
  Phenomena in Massive Stars and Starburst Galaxies, ed. K.~A. van~der Hucht,
  G.~Koenigsberger, \& P.~R.~J. Eenens (San Francisco: Astronomical Society of
  the Pacific), 386

\bibitem[{Prinja {et~al.}(1990)Prinja, Barlow, \& Howarth}]{prinja:1990}
Prinja, R.~K., Barlow, M.~J., \& Howarth, I.~D. 1990, \apj, 361, 607

\bibitem[{Robert {et~al.}(1990)Robert, Moffat, Bastien, St-Louis, \&
  Drissen}]{robert:1990}
Robert, C., Moffat, A.~F.~J., Bastien, P., St-Louis, N., \& Drissen, L. 1990,
  \apj, 359, 211

\bibitem[{{Rodrigues} \& {Magalhaes}(1995)}]{rodrigues:1995}
{Rodrigues}, C.~V. \& {Magalhaes}, A.~M. 1995, in IAU Symp., Vol. 163,
  {Wolf-Rayet Stars: Binaries; Colliding Winds; Evolution}, ed. K.~A. van~der
  Hucht \& P.~M. Williams (Dordrecht: Kluwer), 260

\bibitem[{Schmutz {et~al.}(1989)Schmutz, Hamann, \& Wessolowski}]{schmutz:1989}
Schmutz, W., Hamann, W.-R., \& Wessolowski, U. 1989, \aap, 210, 236

\bibitem[{Shore \& Brown(1988)}]{shore:1988}
Shore, S.~N. \& Brown, D.~N. 1988, \apj, 334, 1021

\bibitem[{St-Louis {et~al.}(1988)St-Louis, Moffat, Drissen, Bastien, \&
  Robert}]{stlouis:1988}
St-Louis, N., Moffat, A.~F.~J., Drissen, L., Bastien, P., \& Robert, C. 1988,
  \apj, 330, 286

\bibitem[{St-Louis {et~al.}(1993)St-Louis, Moffat, Lapointe, Efimov,
  Shakhovskoj, Fox, \& Piirola}]{stlouis:1993}
St-Louis, N., Moffat, A.~F.~J., Lapointe, L., Efimov, Y.~S., Shakhovskoj,
  N.~M., Fox, G.~K., \& Piirola, V. 1993, \apj, 410, 342

\bibitem[{Stevens {et~al.}(1992)Stevens, Blondin, \& Pollock}]{stevens:1992}
Stevens, I.~R., Blondin, J.~M., \& Pollock, A.~M.~T. 1992, \apj, 386, 265

\bibitem[{{Stevens} {et~al.}(1996){Stevens}, {Corcoran}, {Willis}, {Skinner},
  {Pollock}, {Nagase}, \& {Koyama}}]{stevens:1996}
{Stevens}, I.~R., {Corcoran}, M.~F., {Willis}, A.~J., {Skinner}, S.~L.,
  {Pollock}, A.~M.~T., {Nagase}, F., \& {Koyama}, K. 1996, \mnras, 283, 589

\bibitem[{{Stevens} \& {Howarth}(1999)}]{stevens:1999}
{Stevens}, I.~R. \& {Howarth}, I.~D. 1999, \mnras, 302, 549

\bibitem[{Stevens \& Pollock(1994)}]{stevens:1994}
Stevens, I.~R. \& Pollock, A.~M.~T. 1994, \mnras, 269, 226

\bibitem[{Underhill {et~al.}(1990)Underhill, Grieve, \& Louth}]{underhill:1990}
Underhill, A.~B., Grieve, G.~R., \& Louth, H. 1990, \pasp, 102, 749

\bibitem[{Underhill {et~al.}(1988)Underhill, Yang, \& Hill}]{underhill:1988}
Underhill, A.~B., Yang, S., \& Hill, G.~M. 1988, \pasp, 100, 1256

\bibitem[{Usov(1990)}]{usov:1990}
Usov, V.~V. 1990, \apss, 167, 297

\bibitem[{Usov(1992)}]{usov:1992}
---. 1992, \apj, 389, 635

\bibitem[{{van der Hucht}(2001)}]{vanderhucht:2001}
{van der Hucht}, K.~A. 2001, New Astronomy Review, 45, 135

\bibitem[{Vanbeveren {et~al.}(1998)Vanbeveren, De~Loore, \&
  Van~Rensbergen}]{vanbeveren:1998}
Vanbeveren, D., De~Loore, C., \& Van~Rensbergen, W. 1998, \aapr, 9, 63

\bibitem[{{Vishniac}(1983)}]{vishniac:1983}
{Vishniac}, E.~T. 1983, \apj, 274, 152

\bibitem[{{Vishniac}(1994)}]{vishniac:1994}
---. 1994, \apj, 428, 186

\bibitem[{{Wilson}(1939)}]{wilson:1939}
{Wilson}, O.~C. 1939, \pasp, 51, 55

\bibitem[{Wolf {et~al.}(1999)Wolf, Henning, \& Stecklum}]{wolf:1999}
Wolf, S., Henning, T., \& Stecklum, B. 1999, \aap, 349, 839

\bibitem[{Wright \& Barlow(1975)}]{wright:1975}
Wright, A.~E. \& Barlow, M.~J. 1975, \mnras, 170, 41

\end{thebibliography}

\end{document}